\documentclass[prd,twocolumn,floatfix,amsmath,nofootinbib,amssymb,floatfix]{revtex4}
\usepackage{graphicx,color,dcolumn,booktabs,bm}
\usepackage{longtable,lscape}
\usepackage{txfonts}
\usepackage{overpic}
\usepackage{amssymb}
\usepackage{indentfirst}
\usepackage{feynmf}   
\usepackage{slashed}  
\usepackage{cases}
\usepackage{color}
\usepackage{multirow}
\usepackage{epstopdf}
\usepackage{longtable}
\usepackage{graphicx,color,dcolumn,booktabs,bm}
\usepackage[colorlinks,
            citecolor=blue,
            anchorcolor=red,
            menucolor=red,
            linkcolor=red,
            filecolor=red,
            runcolor=red,
            urlcolor=blue,
            frenchlinks=red]{hyperref}

\graphicspath{{Figures/}} %

\allowdisplaybreaks

\begin{document}

\title{Emergence of charm-strange dibaryons with negative parity via baryon-baryon interactions}

\author{Yu-Yue Cui$^{1}$}
\author{Xiao-Mei Tang$^{1}$}
\author{Qi Huang$^{2}$}\email{06289@njnu.edu.cn}
\author{Rui Chen$^{1,3}$}\email{chenrui@hunnu.edu.cn}

\affiliation{
$^1$Key Laboratory of Low-Dimensional Quantum Structures and Quantum Control of Ministry of Education, Department of Physics and Synergetic Innovation Center for Quantum Effects and Applications, Hunan Normal University, Changsha 410081, China\\
$^2$School of Physics and Technology, Nanjing Normal University, Nanjing 210023, China\\
$^3$Hunan Research Center of the Basic Discipline for Quantum Effects and Quantum Technologies, Hunan Normal University, Changsha 410081, China}
\date{\today}

\begin{abstract}
Within the framework of the one-boson-exchange model, we systematically perform a coupled channel analysis of the $P-$wave interactions between a charm baryon and light baryon, the involved channels include $\Xi_cN$, $\Lambda_c\Sigma$, $\Xi_c^{\prime}N$, $\Sigma_c\Lambda$, $\Xi_c^*N$, $\Sigma_c^*\Lambda$, $\Sigma_c\Sigma$, and $\Sigma_c^*\Sigma$. Our results can predict several possible molecular candidates, such as a $\Xi_c^{\prime}N$ molecule with $I(J^P)=0(1^-)$, $\Xi_c^{*}N$ molecules with $0(0^-, 1^-, 2^-)$, $\Sigma_c\Sigma$ molecules with $0(1^-)$ and $1(1^-)$, and $\Sigma_c^*\Sigma$ molecules with $0(0^-, 1^-, 2^-)$, and $1(2^-)$. The coupled channel effects significantly influence the formation of the $\Sigma_c\Sigma$ molecule with $1(1^-)$. Furthermore, we analyze the phase shifts for these coupled channel systems. Our analysis not only confirms the existence of the predicted molecules but also identifies potential resonant dibaryons, including a $\Sigma_c\Sigma$ shape-type resonance with $1(1^-)$, a $\Sigma_c^*\Sigma$ shape-type resonance with $1(0^-)$, a $\Lambda_c\Sigma/\Sigma_c\Sigma$ coupled Feshbach-type resonance with $1(1^-)$, and  $\Lambda_c\Sigma/\Sigma_c^*\Sigma$ coupled Feshbach-type resonances with $1(0^-, 2^-)$.
\end{abstract}


\maketitle

\section{introduction}

In recent decades, experiments have reported a series of new hadron structures, which not only enrich the hadron spectrum, but also provide new insights for understanding the complex non-perturbative behavior in Quantum Chromodynamics (QCD). Among these newly discovered structures, many of them exhibit two key characteristics: first, their properties are inconsistent with the predictions of the conventional quark model, where baryons are three-quark states and mesons are quark-antiquark pairs; second, their masses are very close to some specific threshold of a pair of traditional hadrons, such as the $P_c/P_{cs}$ states, the $T_{cc}$ state, and others. Thus, theorists are inspired to propose various explanations, such as molecular states, compact multiquarks, hybrids, glueballs, hadro-charmonium states, and kinematical effects (see review papers~\cite{Liu:2013waa, Hosaka:2016pey, Chen:2016qju, Richard:2016eis, Lebed:2016hpi, Brambilla:2019esw, Liu:2019zoy, Chen:2022asf, Olsen:2017bmm, Guo:2017jvc, Meng:2022ozq} for more details).

Among the various explanations for new hadron structures, the molecular states have received significant attentions due to its simple physical image and its success in explaining so many exotic states. Under this framework, a deeper understanding of hadron interactions then becomes so important, since it not only can uncover the formation mechanisms of hadronic molecules, but also can predict new states. Compared to hadron interactions involving higher partial waves, $S-$wave interactions often dominate at low energies or long distances, facilitating the formation of stable molecular states. This is why most near-threshold new hadron states are often interpreted as $S-$wave hadronic molecular states. For instance, the $X(3872)$ are often assigned as an isoscalar $D\bar{D}^*$ molecular state with $J^{PC}=1^{++}$ \cite{Wong:2003xk, Swanson:2003tb, Suzuki:2005ha, Liu:2008fh, Thomas:2008ja,Liu:2008tn,Lee:2009hy}, the $P_c$ states are commonly suggested to be  $\Sigma_c\bar{D}^{(*)}$ molecular pentaquarks \cite{Wu:2010jy, Wang:2011rga,Yang:2011wz,Uchino:2015uha,Karliner:2015ina,Chen:2019asm,Liu:2019tjn,Yamaguchi:2019seo,Chen:2019bip,Xiao:2019aya,Meng:2019ilv,PavonValderrama:2019nbk,He:2019ify,Du:2019pij,Burns:2019iih,Wang:2019ato}. In contrast, since higher partial-wave interactions require specific conditions to become significant, there only a few explorations of hadron interactions via $P-$wave interactions have been carried out \cite{Yamaguchi:2011qw, Shimizu:2019jfy, Ohkoda:2011vj, Ohkoda:2012hv, Wang:2024ukc, Lin:2024qcq, Sakai:2025djx, Lu:2025zae, Chen:2025gxe, Yu:2021lmb}.

Obviously, with the accumulation of experimental data, it is now very necessary to investigate the hadrons interactions via the $P-$wave interactions, by which one then can explore the existence of possible hadronic molecular states with different parity than the $S-$wave interactions. For $P-$wave systems, it always involves a repulsive centrifugal force that described by $l(l+1)/2\mu r^2$, which may reduce the likelihood of forming hadronic molecules. However, hadrons actually can be momentarily trapped by the barriers created by these repulsive forces, making it still possible to search for shape-type resonances in $P-$wave interactions. Recently, it is noteworthy that the BESIII Collaboration \cite{BESIII:2024ths} confirmed the existence of a negative parity structure $G(3900)$ reported by BaBar \cite{BaBar:2006qlj, BaBar:2008drv} and Belle \cite{Belle:2007qxm} with high significance, after performed a precise measurement of the $e^++e^-\to D\bar{D}$ Born cross sections. Since its mass is very close to the $D\bar{D}^*/D^*\bar{D}$ threshold, by using a unified meson-exchange model, Lin \textit{et al.} \cite{Lin:2024qcq} found it may be the first $P-$wave $D\bar{D}^*/D^*\bar{D}$ molecular resonance. In addition, their model can simultaneously describe the features of the $\chi_{c1}(3872)$, $Z_c(3900)$ and $T_{cc}(3875)$.

Due to the current experimental status, compared to meson-meson and meson-baryon molecular states, there exists much less works on the di-baryon system. However, as another piece of molecular state, di-baryon system is also worthy to be further studied to examine the unique properties of interactions. Very recently, we studied the interactions between a heavy baryon and a light baryon using the one-boson-exchange (OBE) effective potentials and predicted the existence of possible $\Xi_c^{(\prime,*)}N$ charm-strange molecular dibaryons with positive parity \cite{Huo:2024eew}, which provides a unique opportunity to investigate the $S$-wave dynamics of strong interactions in multi-quark systems. Besides our investigations, there are also several theoretical explorations of the charm-strange dibaryon with positive parity. For example, in Ref.~\cite{Kong:2022rvd}, the authors systematically investigated the dibaryons with charm number $C=1$ and strangeness number $S =\pm 1$ by examining the interactions of charmed and strange baryons using the OBE model, where a series of $S-$wave bound states are found after solving the quasipotential Bethe-Salpeter equation. In Ref.~\cite{Wang:2024riu}, Wang and colleagues investigated the possible molecular states formed by the combinations of hyperons, charmed baryons, and doubly charmed baryons using the effective potentials at the quark level. In Ref.~\cite{Cui:2024zfs}, the authors predicted several bound states or resonance states of singly charmed di-baryon systems with strangeness numbers $S=-1$, $-3$ and $-5$ within the chiral quark model framework.

Although there exists some works on the $S$-wave di-baryon system, the existence of the charm-strange di-baryons with negative parity has been rarely explored. Thus, it is particularly interesting and important to extend our previous investigations and systematically study the $P-$wave interactions between heavy and light baryons. In this work, we continue to use the OBE model, an approach commonly employed to describe hadron interactions \cite{Chen:2015loa, Chen:2016ypj, Chen:2019asm, Chen:2019uvv, Chen:2021vhg, Chen:2020kco, Chen:2020yvq, Chen:2022dad, Liu:2019tjn, He:2019ify, Yamaguchi:2019seo, Burns:2019iih}, to study the $P-$wave interactions for the $\Xi_c^{(\prime,*)}N$, $\Lambda_c\Lambda$, $\Sigma_c^{(*)}\Lambda$, $\Lambda_c\Sigma$, and $\Sigma_c^{(*)}\Sigma$ systems. With the OBE effective potentials obtained, we then search for the solutions to the bound state by solving the coupled channel Schr\"{o}dinger equations. Ultimately, we can predict reasonable molecular candidates.

In addition to predicting molecular states, we analyze the phase shifts for the discussed channels, as phase shifts and hadronic resonances are closely related. Generally, resonances can be categorized as shape-type or Feshbach-type, with coupled-channel effects playing a crucial role in generating Feshbach-type resonances. With these efforts, we can also provide valuable insights into the presence and characteristics of exotic charm-strange resonant dibayons.

This paper is organized as follows. After this introduction, we introduce the effective potentials of the OBE for the systems discussed in Sec.~\ref{sec2}. In Sec.~\ref{sec3}, we present the phase shifts for the systems discussed. The paper ends with a summary in Sec. \ref{sec4}.

\section{System description}\label{sec2}

In this work, we investigate the $P-$wave interactions between a heavy baryon and light baryon by using the OBE model. The general procedures for deducing the OBE effective potentials can be divided into three steps. Firstly, one can write down the scattering amplitudes in $t-$channel based on the relevant effective Lagrangians. Secondly, we can derive the OBE effective potentials in momentum space in a Breit approximation, i.e., $\mathcal{V}_{E}(\bm{q}) = -{\mathcal{M}_E} /{{\sqrt{\prod_i2M_i\prod_f2M_f}}}$. Here, $M_i$ and $M_f$ are the masses of the initial states and final states, respectively. Finally, we can obtain the effective potentials in coordinate space by performing a Fourier transformation, i.e., $\mathcal{V}_{E}({r}) = \int\frac{d^3\bm{q}}{(2\pi)^3}e^{i\bm{q}\cdot\bm{r}}\mathcal{V}_{E}(\bm{q})\mathcal{F}^2(q^2,m_E^2)$, here, $\mathcal{F}^2(q^2,m_E^2)$ is a form factor, which is introduced to compensate the off-shell effects of the exchanged mesons. In this work, we use the monopole-type form factor, $\mathcal{F}(q^2,m_E^2)=(\Lambda^2-m_E^2)/(\Lambda^2-q^2)$, with $\Lambda$, $m_E$, and $q$ being the cutoff, the mass, and four-momentum of the exchanged mesons, respectively. According to the experience of the nucleon-nucleon interactions \cite{Tornqvist:1993ng, Tornqvist:1993vu}, the reasonable cutoff value is taken around 1.00 GeV.

Effective Lagrangians describing the interactions between the heavy baryons and the light mesons are constructed in terms of the heavy quark symmetry, chiral symmetry and hidden local symmetry \cite{Liu:2011xc}, i.e.,
\begin{eqnarray}
\mathcal{L}_{\mathcal{B}_{\bar{3}}} &=& l_B\langle\bar{\mathcal{B}}_{\bar{3}}\sigma\mathcal{B}_{\bar{3}}\rangle
          +i\beta_B\langle\bar{\mathcal{B}}_{\bar{3}}v^{\mu}(\mathcal{V}_{\mu}-\rho_{\mu})\mathcal{B}_{\bar{3}}\rangle,\label{lag1}\\
\mathcal{L}_{\mathcal{B}_{6}} &=&  l_S\langle\bar{\mathcal{S}}_{\mu}\sigma\mathcal{S}^{\mu}\rangle
         -\frac{3}{2}g_1\varepsilon^{\mu\nu\lambda\kappa}v_{\kappa}
         \langle\bar{\mathcal{S}}_{\mu}\mathcal{A}_{\nu}\mathcal{S}_{\lambda}\rangle\nonumber\\
        &&+i\beta_{S}\langle\bar{\mathcal{S}}_{\mu}v_{\alpha}\left(\mathcal{V}^{\alpha}-\rho^{\alpha}\right) \mathcal{S}^{\mu}\rangle
         +\lambda_S\langle\bar{\mathcal{S}}_{\mu}F^{\mu\nu}(\rho)\mathcal{S}_{\nu}\rangle,\label{lag2}\\
\mathcal{L}_{\mathcal{B}_{\bar{3}}\mathcal{B}_6} &=& ig_4\langle\bar{\mathcal{S}^{\mu}}\mathcal{A}_{\mu}\mathcal{B}_{\bar{3}}\rangle+i\lambda_I\varepsilon^{\mu\nu\lambda\kappa}v_{\mu}\langle \bar{\mathcal{S}}_{\nu}F_{\lambda\kappa}\mathcal{B}_{\bar{3}}\rangle+h.c..\label{lag03}
\end{eqnarray}
Here, $v=(1,\textbf{0})$ is the four velocity, $\mathcal{A}_{\mu}$ and $\mathcal{V}_{\mu}$ are the axial current and vector current, respectively, which have the forms of $\mathcal{A}_{\mu}=\frac{1}{2}(\xi^{\dag}\partial_{\mu}\xi-\xi\partial_{\mu}\xi^{\dag})$ and $\mathcal{V}_{\mu}=\frac{1}{2}(\xi^{\dag}\partial_{\mu}\xi+\xi\partial_{\mu}\xi^{\dag})$, with $\xi=\text{exp}(i{P}/f_{\pi})$ and $f_{\pi}=132$ MeV. $F^{\mu\nu}(\rho)=\partial^{\mu}\rho^{\nu}-\partial^{\nu}\rho^{\mu}
+\left[\rho^{\mu},\rho^{\nu}\right]$, with $\rho^{\mu}=ig_V{V}^{\mu}/\sqrt{2}$. The superfield $\mathcal{S}_{\mu} =-\sqrt{\frac{1}{3}}(\gamma_{\mu}+v_{\mu})\gamma^5\mathcal{B}_6+\mathcal{B}_{6\mu}^*$ is expressed as a combination of heavy baryon fields $\mathcal{B}_6$ with $J^P=1/2^+$ and $\mathcal{B}^*_6$ with $J^P=3/2^+$.

For the interactions between the light baryons and the light mesons, the relevant effective Lagrangians are constructed in $SU(3)$ symmetry, which read as
\begin{eqnarray}
\mathcal{L}_{B B \sigma} &  = & g_{B B \sigma} \bar{B} \sigma B, \\
\mathcal{L}_{B B P} & = & \frac{g_{B B P}}{m_{P}} \bar{B} \gamma^{5} \gamma^{\mu} \partial_{\mu} P B, \\
\mathcal{L}_{B B V} & = & g_{B B V} \bar{B}\gamma^{\mu}V_{\mu}B-\frac{f_{B B V}}{2 m_{B}}\bar{B} \sigma^{\mu v} \partial_{v} V_{\mu} B.
\end{eqnarray}
Here, the baryon and meson multiplets $\mathcal{B}_{\bar{3}}$, $\mathcal{B}_6$, ${P}$, ${V}$, and $B$ are explicitly written as
\begin{eqnarray}
\mathcal{B}_{\bar{3}} &=& \left(\begin{array}{ccc}
        0    &\Lambda_c^+      &\Xi_c^+\\
        -\Lambda_c^+       &0      &\Xi_c^0\\
        -\Xi_c^+      &-\Xi_c^0     &0
\end{array}\right),\\
\mathcal{B}_6^{(*)} &=& \left(\begin{array}{ccc}
         \Sigma_c^{{(*)}++}                  &\frac{\Sigma_c^{{(*)}+}}{\sqrt{2}}     &\frac{\Xi_c^{(',*)+}}{\sqrt{2}}\\
         \frac{\Sigma_c^{{(*)}+}}{\sqrt{2}}      &\Sigma_c^{{(*)}0}    &\frac{\Xi_c^{(',*)0}}{\sqrt{2}}\\
         \frac{\Xi_c^{(',*)+}}{\sqrt{2}}    &\frac{\Xi_c^{(',*)0}}{\sqrt{2}}      &\Omega_c^{(*)0}
\end{array}\right),\\
P &=& \left(\begin{array}{ccc}
\frac{\pi^0}{\sqrt{2}}+\frac{\eta}{\sqrt{6}} &\pi^+ &K^+ \\
\pi^- &-\frac{\pi^0}{\sqrt{2}}+\frac{\eta}{\sqrt{6}} &K^0 \\
K^- &\bar{K}^0 &-\sqrt{\frac{2}{3}}\eta
\end{array}\right),\\
V &=& \left(\begin{array}{ccc}
\frac{\rho^0}{\sqrt{2}}+\frac{\omega}{\sqrt{2}} &\rho^+ &K^{*+} \nonumber\\
\rho^- &-\frac{\rho^0}{\sqrt{2}}+\frac{\omega}{\sqrt{2}} &K^{*0} \nonumber\\
K^{*-} &\bar{K}^{*0} &\phi
\end{array}\right),\\
B &=& \left(\begin{array}{ccc}
\frac{\Sigma^0}{\sqrt{2}}+\frac{\Lambda}{\sqrt{6}}    &\Sigma^+    &p\\
\Sigma^-  &-\frac{\Sigma^0}{\sqrt{2}}+\frac{\Lambda}{\sqrt{6}}   &n\\
\Xi^-      &\Xi^0     &-\frac{2}{\sqrt{6}}\Lambda\end{array}\right),
\end{eqnarray}
respectively. With the help of the quark model, we can obtain the relations between the coupling constants for the heavy baryons and the nucleon-nucleon interactions~\cite{Liu:2011xc}, i.e., $l_S=-2l_B=-\frac{2}{3}g_{\sigma NN}$, $g_1=\frac{2\sqrt{2}}{3}g_4=-\frac{2\sqrt{2}f_{\pi}g_{\pi NN}}{5 M_N}$, $\beta_Sg_V=-2\beta_Bg_V=-4g_{\rho NN}$, $\lambda_Sg_V=-\sqrt{8}\lambda_Ig_V=-\frac{6(g_{\rho NN}+f_{\rho NN})}{5M_N}$. The coupling constants for the nucleon-nucleon interactions are given in Refs. \cite{Machleidt:2000ge,Machleidt:1987hj,Cao:2010km}. In Table \ref{constants}, we collect the values of all the coupling constants.
\renewcommand\tabcolsep{0.1cm}
\renewcommand{\arraystretch}{1.7}
\begin{table}[!htbp]
\caption{Coupling constants adopted in our calculations.}\label{constants}
{\begin{tabular}{cccccc}
\toprule[1pt]
\toprule[1pt]
$\frac{g^2_{\sigma NN}}{4\pi}=5.69$   &$g_{NN\eta}=0.33$   &$\frac{g^2_{\pi NN}}{4\pi}=0.07$   &$g_{\Lambda\Lambda\omega}=7.98$\\   
$\frac{g^2_{\rho NN}}{4\pi}=0.81$ &$f_{\Lambda\Lambda\omega}=-9.73$
&$\frac{f_{\rho NN}}{g_{\rho NN}}=6.10$ &$g_{\Lambda\Lambda\eta}=-0.67$\\
$\frac{g_{\omega NN}^2}{4\pi}=20.00$ &$g_{\Lambda NK^*}=-6.08$
&$\frac{f_{\omega NN}}{g_{\omega NN}}=0.00$   &$f_{\Lambda NK^*}=-16.85$\\
$f_{\Lambda\Sigma\rho}=16.85$  &$g_{\Lambda\Sigma\rho}=-0.55$
&$g_{\Lambda\Sigma\pi}=0.67$   &$g_{N \Sigma K}=0.19$\\
$g_{ \Sigma\Sigma \rho/\omega}=7.34$  &$f_{\Sigma\Sigma \rho/\omega}=9.73$
&$g_{\Sigma\Sigma \pi}=0.77$   &$g_{\Sigma\Sigma \eta}=0.67$\\
$g_{ N\Sigma K^*}=-4.15$     &$f_{ N\Sigma K^*}=9.73$
&$g_{ \Lambda N K}=-1.00$\\
\bottomrule[1pt]\bottomrule[1pt]
\end{tabular}}
\end{table}

\renewcommand\tabcolsep{0.5cm}
\renewcommand{\arraystretch}{1.7}
\begin{table*}[!htbp]\center
  \caption{Possible channels involved in our calculation. Here, the first column contains the spin-parity quantum numbers corresponding to the channels. }\label{channels}
  {\begin{tabular}{c|llllllll}
\toprule[1pt]
  $I(J^P)$\quad   &\multicolumn{5}{c}{Channels}\\
\hline
$0(0^-)$     &$\Xi_cN:|{}^3P_{0}\rangle$
 &$\Xi_c^{\prime}N:|{}^3P_{0}\rangle$
  &$\Xi_c^{*}N:|{}^3P_{0}\rangle$
   &$\Sigma_c\Sigma:|{}^3P_{0}\rangle$
    &$\Sigma_c^{*}\Sigma:|{}^3P_{0}\rangle$
                   \\
$0(1^-)$    &$\Xi_cN:|{}^1P_{1}/{}^3P_{1}\rangle$
 &$\Xi_c^{\prime}N:|{}^1P_{1}{}/^3P_{1}\rangle$
  &$\Xi_c^{*}N:|{}^3P_{1}/{}^5P_{1}\rangle$
   &$\Sigma_c\Sigma:|{}^1P_{1}/{}^3P_{1}\rangle$
    &$\Sigma_c^{*}\Sigma:|{}^3P_{1}/{}^5P_{1}\rangle$\\
$0(2^-)$    &$\Xi_cN:|{}^3P_{2}\rangle$
 &$\Xi_c^{\prime}N:|{}^3P_{2}\rangle$
  &$\Xi_c^{*}N:|{}^3P_{2}/{}^5P_{2}\rangle$
   &$\Sigma_c\Sigma:|{}^3P_{2}\rangle$
    &$\Sigma_c^{*}\Sigma:|{}^3P_{2}/{}^5P_{2}\rangle$\\
$1(0^-)$     &$\Xi_cN:|{}^3P_{0}\rangle$
    &$\Lambda_c\Sigma:|{}^3P_{0}\rangle$
 &$\Xi_c^{\prime}N:|{}^3P_{0}\rangle$
 &$\Sigma_c\Lambda:|{}^3P_{0}\rangle$
  &$\Xi_c^{*}N:|{}^3P_{0}\rangle$\\
  &$\Sigma_c^{*}\Lambda:|{}^3P_{0}\rangle$
   &$\Sigma_c\Sigma:|{}^3P_{0}\rangle$
    &$\Sigma_c^{*}\Sigma:|{}^3P_{0}\rangle$\\
$1(1^-)$     &$\Xi_cN:|{}^1P_{1}/{}^3P_{1}\rangle$
    &$\Lambda_c\Sigma:|{}^1P_{1}/{}^3P_{1}\rangle$
 &$\Xi_c^{\prime}N:|{}^1P_{1}/{}^3P_{1}\rangle$\\
 &$\Sigma_c\Lambda:|{}^1P_{1}/{}^3P_{1}\rangle$
  &$\Xi_c^{*}N:|{}^3P_{1}/{}^5P_{1}\rangle$
  &$\Sigma_c^{*}\Lambda:|{}^3P_{1}/{}^5P_{1}\rangle$
   &$\Sigma_c\Sigma:|{}^1P_{1}/{}^3P_{1}\rangle$
    &$\Sigma_c^{*}\Sigma:|{}^3P_{1}/{}^5P_{1}\rangle$\\
$1(2^-)$     &$\Xi_cN:|{}^3P_{2}\rangle$
    &$\Lambda_c\Sigma:|{}^3P_{2}\rangle$
 &$\Xi_c^{\prime}N:|{}^3P_{2}\rangle$
 &$\Sigma_c\Lambda:|{}^3P_{2}\rangle$
  &$\Xi_c^{*}N:|{}^3P_{2}/{}^5P_{2}\rangle$\\
  &$\Sigma_c^{*}\Lambda:|{}^3P_{2}/{}^5P_{2}\rangle$
   &$\Sigma_c\Sigma:|{}^3P_{2}\rangle$
    &$\Sigma_c^{*}\Sigma:|{}^3P_{2}/{}^5P_{2}\rangle$\\
\bottomrule[1pt]
\end{tabular}}
\end{table*}

In this work, the discussed systems include $\Xi_cN$, $\Lambda_c\Sigma$, $\Xi_c^{\prime}N$, $\Sigma_c\Lambda$, $\Xi_c^*N$, $\Sigma_c^*\Lambda$, $\Sigma_c\Sigma$, $\Sigma_c^*\Sigma$ channels. In Table \ref{channels}, we list the channels with different quantum number configurations, where $|{}^{2S+1}L_J\rangle$ are spin-orbit wave functions that have the general form as
\begin{eqnarray}
\mathcal{B}_c(1/2)B: && \sum_{m,n,m_L}C_{\frac{1}{2},m;\frac{1}{2},n}^{S,ms}C_{S,m_S;L,m_L}^{J,m_J}
       \chi_{\frac{1}{2},m}\chi_{\frac{1}{2},n}Y_{L,m_L},\\
\mathcal{B}_c(3/2)B: && \sum_{m,n,m_L}C_{\frac{1}{2},m;\frac{3}{2},n}^{S,m_S}C_{S,m_S;L,m_L}^{J,m_J}
       \chi_{\frac{1}{2},m}\Phi_{\frac{3}{2},n}Y_{L,m_L}.
\end{eqnarray}
Here, $\mathcal{B}_c(1/2,3/2)$ denote the heavy baryons with spin $J=1/2$ and $3/2$, respectively. $B$ stands for the light baryons. $C_{\frac{1}{2},m;\frac{1}{2},n}^{S,ms}$, $C_{\frac{1}{2},m;\frac{3}{2},n}^{S,m_S}$, and $C_{S,m_S;L,m_L}^{J,m_J}$ are the Clebsch-Gordan coefficients. $\chi_{\frac{1}{2},m}$ and $Y_{L,m_L}$ denote the spin wave function for the baryon with $J=1/2$ and the spherical harmonics function, respectively. $\Phi_{\frac{3}{2},n}=\sum_{n_1,n_2}\langle\frac{1}{2},n_1;1,n_2|\frac{3}{2},n\rangle
\chi_{\frac{1}{2},n_1}\epsilon^{n_2}$ is the spin wave function for the baryon with $J=3/2$, where $\epsilon$ stands for the polarization vector, $\epsilon_{\pm1}= \frac{1}{\sqrt{2}}\left(0,\pm1,i,0\right)$ and $\epsilon_{0} =\left(0,0,0,-1\right)$.

The flavor wave functions $|I,I_3\rangle$ for all the investigated systems can be expressed as
\begin{eqnarray}
\Xi_c^{(\prime,*)}N: &&\left\{\begin{array}{l}|0,0\rangle=(|\Xi_c^{(',*)+}n\rangle-|\Xi_c^{(',*)0}p\rangle)/\sqrt{2},\\
        |1,1\rangle=|\Xi_c^{(',*)+}p\rangle,\\
        |1,0\rangle=(|\Xi_c^{(',*)+}n\rangle+|\Xi_c^{(',*)0}p\rangle)/\sqrt{2},\\
        |1,-1\rangle=|\Xi_c^{(',*)0}n\rangle,\end{array}\right.\\
\Lambda_c\Sigma: &&\left\{\begin{array}{l}
|1,1\rangle=|\Lambda_c^+\Sigma^+\rangle,\\
        |1,0\rangle=|\Lambda_c^+\Sigma^0\rangle,\\
        |1,-1\rangle=|\Lambda_c^+\Sigma^-\rangle,\end{array}\right.\\
\Sigma_c^{(*)}\Lambda: &&\left\{\begin{array}{l}|1,1\rangle=|\Sigma_c^{(*)++}\Lambda^0\rangle,\\
        |1,0\rangle=|\Sigma_c^{(*)+}\Lambda^0\rangle,\\
        |1,-1\rangle=|\Sigma_c^{(*)0}\Lambda^0\rangle,\\
        \end{array}\right.,\\
\Sigma_c^{(*)}\Sigma: &&\left\{\begin{array}{l}|0,0\rangle=(|\Sigma_c^{(*)++}\Sigma^-\rangle+|\Sigma_c^{(*)0}\Sigma^+\rangle)/\sqrt{
3},\\
        |1,1\rangle=(|\Sigma_c^{(*)++}\Sigma^0\rangle-|\Sigma_c^{(*)+}\Sigma^+\rangle)/\sqrt{
2},\\
        |1,0\rangle=(|\Sigma_c^{(*)++}\Sigma^-\rangle-|\Sigma_c^{(*)0}\Sigma^+\rangle)/\sqrt{
2},\\
        |1,-1\rangle=(|\Sigma_c^{(*)+}\Sigma^-\rangle-|\Sigma_c^{(*)0}\Sigma^0\rangle)/\sqrt{
2}.\end{array}\right.
\end{eqnarray}

\renewcommand\tabcolsep{0.1cm}
\renewcommand{\arraystretch}{1.7}

\begin{table*}[!htbp]
\caption{A summary of the total OBE effective potentials for all the discussed processes. Variables in the effective potentials are defined as $\Lambda_i^2=\Lambda^2-q_i^2$, $m_i^2=m^2-q_i^2$ with $i=0,1,\ldots$ And $q_i$ is zero-component of the momentum of the exchange mesons, which equals $q_i== \frac{m_{1}^2-m_2^2+m_4^2-m_{3}^2}{2m_{3}+2m_4}$ for the $h_1h_2\to h_3h_4$ process.}\label{obe}
{\begin{tabular}{c|lll|lll}
\toprule[1pt]
\toprule[1pt]
Type     &Processes    &Isospin    &Total effective potentials    &Processes    &Isospin    &Total effective potentials\\\hline
${\mathcal{B}_{\bar{3}}B\to\mathcal{B}_{\bar{3}}B}$
  &$\Xi_cN$$\to$$\Xi_cN$  &$I=0$    &$2\mathcal{V}_{\sigma}^a+\frac{1}{\sqrt{2}}\mathcal{V}_{\omega}^a-\frac{3}{\sqrt{2}}\mathcal{V}_{\rho}^a$
  & $\Xi_cN\to\Lambda_c\Sigma$    &$I=1$    &$-\sqrt{2}\mathcal{V}_{K0}^a-\sqrt{2}\mathcal{V}_{K^{*}0}^a$  \\
  &   &$I=1$   &$2\mathcal{V}_{\sigma}^a+\frac{1}{\sqrt{2}}\mathcal{V}_{\omega}^a+\frac{1}{\sqrt{2}}\mathcal{V}_{\rho}^a$  
& $\Lambda_c\Sigma\to\Lambda_c\Sigma$    &$I=1$    &$2\mathcal{V}_{\sigma}^a+\sqrt{2}\mathcal{V}_{\omega}^a$\\\hline
${\mathcal{B}_{\bar{3}}B\to\mathcal{B}_{\bar{6}}B}$
  &$\Xi_cN\to\Xi_c^{\prime}N$  &$I=0$    &$ \frac{\sqrt{3}}{2}\mathcal{V}_{\eta1}^b+ \frac{1}{2}\mathcal{V}_{\omega1}^b- \frac{3}{2}\mathcal{V}_{\pi1}^b- \frac{3}{2}\mathcal{V}_{\rho1}^b$
   &$\Xi_cN\to\Sigma_c\Sigma$  &$I=0$    &$-\sqrt{3}\mathcal{V}_{K2}^b-\sqrt{3}\mathcal{V}_{K^{*}2}^b$         
 \\
&    &$I=1$    &$ \frac{\sqrt{3}}{2}\mathcal{V}_{\eta1}^b+ \frac{1}{2}\mathcal{V}_{\omega1}^b+\frac{1}{2}\mathcal{V}_{\pi1}^b+\frac{1}{2}\mathcal{V}_{\rho1}^b$
&&$I=1$     &$-\sqrt{2}\mathcal{V}_{K2}^b-\sqrt{2}\mathcal{V}_{K^{*}2}^b$           \\
 &$\Xi_cN$$\to$$\Sigma_c\Lambda$  &$I=1$   &$-\mathcal{V}_{K3}^b-
 \mathcal{V}_{K^*3}^b$
 &$\Lambda_c\Sigma\to\Xi_c^{\prime}N$
    &$I=1$  &$\mathcal{V}_{K4}^b+\mathcal{V}_{K^{*}4}^b $\\
     &$\Lambda_c\Sigma\to\Sigma_c\Lambda$
    &$I=1$  &$\mathcal{V}_{\pi5}^b+
 \mathcal{V}_{\rho5}^b$
 &$\Lambda_c\Sigma\to\Sigma_c\Sigma$
    &$I=1$  &$-\sqrt{2}\mathcal{V}_{\pi6}^b$$-\sqrt{2}\mathcal{V}_{\rho6}^b$ \\\hline
${\mathcal{B}_{\bar{3}}B\to\mathcal{B}_{\bar{6}}^*B}$
  &$\Xi_cN$$\to$$\Xi_c^{*}N$  &$I=0$    &$ \frac{\sqrt{3}}{2}\mathcal{V}_{\eta7}^c+ \frac{1}{2}\mathcal{V}_{\omega7}^c- \frac{3}{2}\mathcal{V}_{\pi7}^c- \frac{3}{2}\mathcal{V}_{\rho7}^c$
  &$\Xi_cN\to\Sigma_c^{*}\Sigma$   &$I=0$    &$-\sqrt{3}\mathcal{V}_{K8}^c-\sqrt{3}\mathcal{V}_{K^{*}8}^c$           \\
&    &$I=1$    &$ \frac{\sqrt{3}}{2}\mathcal{V}_{\eta7}^c+ \frac{1}{2}\mathcal{V}_{\omega7}^c+\frac{1}{2}\mathcal{V}_{\pi7}^c+ \frac{1}{2}\mathcal{V}_{\rho7}^c$
&   &$I=1$    &$-\sqrt{2}\mathcal{V}_{K8}^c-\sqrt{2}\mathcal{V}_{K^{*}8}^c$           \\
  &$\Xi_cN\to\Sigma_c^{*}\Lambda$   &$I=1$    &$-\mathcal{V}_{K9}^c-
 \mathcal{V}_{K^*9}^c$
   &$\Lambda_c\Sigma$$\to$$\Xi_c^{*}N$  &$I=1$    &$\mathcal{V}_{K10}^c+\mathcal{V}_{K^{*}10}^c $\\

  &$\Lambda_c\Sigma\to\Sigma_c^{*}\Lambda$   &$I=1$    &$\mathcal{V}_{\pi11}^b$$+
 \mathcal{V}_{\rho11}^b$
  &$\Lambda_c\Sigma\to\Sigma_c^{*}\Sigma$   &$I=1$    & $-\sqrt{2}\mathcal{V}_{\pi12}^b$$-\sqrt{2}\mathcal{V}_{\rho12}^b$         \\
           \hline
${\mathcal{B}_{\bar{6}}B\to\mathcal{B}_{\bar{6}}B}$  &$\Xi_c^{\prime}N\to\Xi_c^{\prime}N$ &$I=0$   &$\mathcal{V}_{\sigma}^d-\frac{1}{2\sqrt{6}}\mathcal{V}_{\eta}^d+\frac{1}{2\sqrt{2}}\mathcal{V}_{\omega}^d-\frac{3}{2\sqrt{2}}\mathcal{V}_{\pi}^d-\frac{3}{2\sqrt{2}}\mathcal{V}_{\rho}^d$
&$\Xi_c^{\prime}N\to\Sigma_c\Sigma$ &$I=0$   &$\sqrt{\frac{3}{2}}\mathcal{V}_{K13}+\sqrt{\frac{3}{2}}\mathcal{V}_{K^{*}13}$

\\
&&$I=1$&$\mathcal{V}_{\sigma}-\frac{1}{2\sqrt{6}}\mathcal{V}_{\eta}+\frac{1}{2\sqrt{2}}\mathcal{V}_{\omega}+\frac{1}{2\sqrt{2}}\mathcal{V}_{\pi}+\frac{1}{2\sqrt{2}}\mathcal{V}_{\rho}$
&&$I=1$   &$\mathcal{V}_{K13}+\mathcal{V}_{K^{*}13}$
\\
&$\Xi_c^{\prime}N\to\Sigma_c\Lambda$ &$I=1$   &$\frac{1}{\sqrt{2}}\mathcal{V}_{K14}+\frac{1}{\sqrt{2}}\mathcal{V}_{K^{*}14}$
&$\Sigma_c\Lambda\to\Sigma_c\Lambda$     &$I=1$
&$\mathcal{V}_{\sigma}+\frac{1}{\sqrt{6}}\mathcal{V}_{\eta}^d+\frac{1}{\sqrt{2}}\mathcal{V}_{\omega}^d$\\
&$\Sigma_c\Sigma\to\Sigma_c\Sigma$
&$I=0$   &$\mathcal{V}_{\sigma}+\frac{1}{\sqrt{6}}\mathcal{V}_{\eta}+\frac{1}{\sqrt{2}}\mathcal{V}_{\omega}-\sqrt{2}\mathcal{V}_{\pi}-\sqrt{2}\mathcal{V}_{\rho}$ &$\Sigma_c\Lambda\to\Sigma_c\Sigma$     &$I=1$
&$\mathcal{V}_{\pi15}+\mathcal{V}_{\rho15}$ \\
&&$I=1$   &$\mathcal{V}_{\sigma}+\frac{1}{\sqrt{6}}\mathcal{V}_{\eta}+\frac{1}{\sqrt{2}}\mathcal{V}_{\omega}-\frac{1}{\sqrt{2}}\mathcal{V}_{\pi}-\frac{1}{\sqrt{2}}\mathcal{V}_{\rho}$ \\\hline
${\mathcal{B}_{\bar{6}}B\to\mathcal{B}_{\bar{6}}^*B}$
&$\Xi_c^{\prime}N\to\Xi_c^*N$     &$I=0$
 &$-\frac{1}{2\sqrt{6}}\mathcal{V}_{\eta16}+\frac{1}{2\sqrt{2}}\mathcal{V}_{\omega16}-\frac{3}{2\sqrt{2}}\mathcal{V}_{\pi16}-\frac{3}{2\sqrt{2}}\mathcal{V}_{\rho16}$
 &$\Xi_c^{\prime}N\to\Sigma_c^*\Sigma$     &$I=0$   &$\sqrt{\frac{3}{2}}\mathcal{V}_{K17}+\sqrt{\frac{3}{2}}\mathcal{V}_{K^{*}17}$
\\
&&$I=1$ &$-\frac{1}{2\sqrt{6}}\mathcal{V}_{\eta35}+\frac{1}{2\sqrt{2}}\mathcal{V}_{\omega16}+\frac{1}{2\sqrt{2}}\mathcal{V}_{\pi16}+\frac{1}{2\sqrt{2}}\mathcal{V}_{\rho16}$
&    &$I=1$  &$\mathcal{V}_{K17}+\mathcal{V}_{K^{*}17}$
\\
&$\Xi_c^{\prime}N\to\Sigma_c^*\Lambda$     &$I=1$  &$\frac{1}{\sqrt{2}}\mathcal{V}_{K18}+\frac{1}{\sqrt{2}}\mathcal{V}_{K^{*}18}$

&$\Sigma_c\Lambda\to\Xi_c^*N$     &$I=1$
&$\frac{1}{\sqrt{2}}\mathcal{V}_{K19}^d+\frac{1}{\sqrt{2}}\mathcal{V}_{K^*19}^d$\\

&$\Sigma_c\Sigma\to\Sigma_c^*\Sigma$     &$I=0$  &$\frac{1}{\sqrt{6}}\mathcal{V}_{\eta20}+\frac{1}{\sqrt{2}}\mathcal{V}_{\omega20}-\sqrt{2}\mathcal{V}_{\pi20}-\sqrt{2}\mathcal{V}_{\rho20}$ &$\Sigma_c\Lambda\to\Sigma_c^*\Lambda$     &$I=1$  &$\frac{1}{\sqrt{6}}\mathcal{V}_{\eta21}^d+\frac{1}{\sqrt{2}}\mathcal{V}_{\omega21}^d$
\\
&&$I=1$   &$\frac{1}{\sqrt{6}}\mathcal{V}_{\eta20}+\frac{1}{\sqrt{2}}\mathcal{V}_{\omega20}-\frac{1}{\sqrt{2}}\mathcal{V}_{\pi20}-\frac{1}{\sqrt{2}}\mathcal{V}_{\rho20}$ &$\Sigma_c\Lambda\to\Sigma_c^*\Sigma$     &$I=1$   &$\mathcal{V}_{\pi22}^d+
 \mathcal{V}_{\rho22}^d$
\\\hline
${\mathcal{B}_{\bar{6}}^*B\to\mathcal{B}_{\bar{6}}B}$&$\Xi_c^{*}N\to\Sigma_c\Sigma$     &$I=0$
 &$\sqrt{\frac{3}{2}}\mathcal{V}_{K23}+\sqrt{\frac{3}{2}}\mathcal{V}_{K^{*}23}$
&$\Sigma_c^{*}\Lambda\to\Sigma_c\Sigma$     &$I=1$
 &$\mathcal{V}_{\pi24}^e$$+\mathcal{V}_{\rho24}^e$
\\
&&$I=1$
 &$\mathcal{V}_{K23}$+$\mathcal{V}_{K^{*}23}$
\\
\hline
${\mathcal{B}_{\bar{6}}^*B\to\mathcal{B}_{\bar{6}}^*B}$&$\Xi_c^{*}N\to\Xi_c^*N$     &$I=0$
 &$\mathcal{V}_{\sigma}-\frac{1}{2\sqrt{6}}\mathcal{V}_{\eta}+\frac{1}{2\sqrt{2}}\mathcal{V}_{\omega}-\frac{3}{2\sqrt{2}}\mathcal{V}_{\pi}-\frac{3}{2\sqrt{2}}\mathcal{V}_{\rho}$
  &$\Xi_c^{*}N\to\Sigma_c^*\Sigma$     &$I=0$
   &$\sqrt{\frac{3}{2}}\mathcal{V}_{K25}+\sqrt{\frac{3}{2}}\mathcal{V}_{K^{*}25}$
\\
&&$I=1$&$\mathcal{V}_{\sigma}-\frac{1}{2\sqrt{6}}\mathcal{V}_{\eta}+\frac{1}{2\sqrt{2}}\mathcal{V}_{\omega}+\frac{1}{2\sqrt{2}}\mathcal{V}_{\pi}+\frac{1}{2\sqrt{2}}\mathcal{V}_{\rho}$
&    &$I=1$  &$\mathcal{V}_{K25}$+$\mathcal{V}_{K^{*}25}$
\\
&$\Xi_c^{*}N\to\Sigma_c^*\Lambda$     &$I=1$
   &$\frac{1}{\sqrt{2}}\mathcal{V}_{K26}$+$\frac{1}{\sqrt{2}}\mathcal{V}_{K^{*}26}$
&$\Sigma_c^*\Lambda\to\Sigma_c^*\Lambda$     &$I=1$  &$\mathcal{V}_{\sigma}+\frac{1}{\sqrt{6}}\mathcal{V}_{\eta}+\frac{1}{\sqrt{2}}\mathcal{V}_{\omega}$
\\
&$\Sigma_c^*\Sigma\to\Sigma_c^*\Sigma$     &$I=0$
 &$\mathcal{V}_{\sigma}+\frac{1}{\sqrt{6}}\mathcal{V}_{\eta}+\frac{1}{\sqrt{2}}\mathcal{V}_{\omega}-\sqrt{2}\mathcal{V}_{\pi}-\sqrt{2}\mathcal{V}_{\rho}$ &$\Sigma_c^*\Lambda\to\Sigma_c^*\Sigma$     &$I=1$  &$\mathcal{V}_{\pi27}+\mathcal{V}_{\rho27}$
\\
&&$I=1$   &$\mathcal{V}_{\sigma}+\frac{1}{\sqrt{6}}\mathcal{V}_{\eta}+\frac{1}{\sqrt{2}}\mathcal{V}_{\omega}-\frac{1}{\sqrt{2}}\mathcal{V}_{\pi}-\frac{1}{\sqrt{2}}\mathcal{V}_{\rho}$ \\
\bottomrule[1pt]\bottomrule[1pt]
\end{tabular}}
\end{table*}

With the above preparations, one can then deduce the concrete OBE effective potentials for all of the investigated processes, which have been collected into Table \ref{obe}. The flavor independent sub-potentials are
\begin{eqnarray}
V_{\sigma}^a &=& -g_{\sigma BB}l_BY(\Lambda,m_{\sigma},r),\\
V_{\rho}^a &=& \frac{g_{BBV}\beta_Bg_V}{\sqrt{2}}
  Y(\Lambda,m_{V},r) \nonumber\\
  &&+\frac{f_{BBV}\beta_Bg_V}{8\sqrt{2}m_B}\left(\frac{1}{m_1}+\frac{1}{m_3}\right)\mathcal{O}_{r}
     Y(\Lambda,m_{V},r),
\end{eqnarray}
for process ${\mathcal{B}_{\bar{3}}B\to\mathcal{B}_{\bar{3}}B}$,
\begin{eqnarray}
V_{P}^b &=&- \frac{g_{BBP}g_4}{3\sqrt{3}f_{\pi}m_P}(\mathcal{A}_{6}\mathcal{O}_{r}+\mathcal{A}_{7}\mathcal{P}_{r})
  Y(\Lambda,m_{P},r),\\
V_{V}^b &=&- \frac{\lambda_Ig_V}{3\sqrt{6}}\left(\frac{g_{BBV}}{2m_1}+\frac{g_{BBV}}{2m_3}+\frac{f_{BBV}}{m_B}\right)(2\mathcal{A}_{6}\mathcal{O}_{r}-\mathcal{A}_{7}\mathcal{P}_{r})\nonumber\\
&&\times Y(\Lambda,m_{V},r),
\end{eqnarray}
for process ${\mathcal{B}_{\bar{3}}B\to\mathcal{B}_{6}B}$,
\begin{eqnarray}
V_{P}^c &=& \frac{g_{BBP}g_4}{3f_{\pi}m_P}(\mathcal{A}_{6}\mathcal{O}_{r}+\mathcal{A}_{7}\mathcal{P}_{r})
  Y(\Lambda,m_{P},r),\\
V_{V}^c &=& \frac{\lambda_Ig_V}{3\sqrt{2}}\left(\frac{g_{BBV}}{2m_1}+\frac{g_{BBV}}{2m_3}+\frac{f_{BBV}}{m_B}\right)(2\mathcal{A}_{6}\mathcal{O}_{r}-\mathcal{A}_{7}\mathcal{P}_{r})\nonumber\\
  &&\times Y(\Lambda,m_{V},r),
\end{eqnarray}
for process ${\mathcal{B}_{\bar{3}}B\to\mathcal{B}_{6}^*B}$,
\begin{eqnarray}
V_{\sigma}^d &=& g_{\sigma BB}l_SY(\Lambda,m_{\sigma},r),\\
V_{P}^d &=& -\frac{g_{BBP}g_1}{3f_{\pi}m_P}(\mathcal{A}_{1}\mathcal{O}_{r}+\mathcal{A}_{2}\mathcal{P}_{r})
 Y(\Lambda,m_{P},r),\\
V_{V}^d &=& -\left(\frac{\beta_Sg_Vf_{BBV}}{8\sqrt{2}m_B}\left(\frac{1}{m_1}+\frac{1}{m_3}\right)+\frac{\lambda_Sg_Vg_{BBV}}{6\sqrt{2} }\left(\frac{1}{m_2}+\frac{1}{m_4}\right)\right)\nonumber\\
&&\times \mathcal{O}_{r}Y(\Lambda,m_{V},r)-\frac{\beta_Sg_Vg_{BBV}}{\sqrt{2}}Y(\Lambda,m_{V},r)\nonumber\\
  &&-\frac{\lambda_Sg_V}{9\sqrt{2} }\left(\frac{g_{BBV}}{2m_1}+\frac{g_{BBV}}{2m_3}+\frac{f_{BBV}}{m_B}\right)(\mathcal{A}_{1}\mathcal{O}_{r}+\mathcal{A}_{2}\mathcal{P}_{r})\nonumber\\
  &&\times Y(\Lambda,m_{V},r),
\end{eqnarray}
for process ${\mathcal{B}_{6}B\to\mathcal{B}_{6}B}$,
\begin{eqnarray}
V_{P}^e &=& +\frac{g_{BBP}g_1}{2\sqrt{3}m_Pf_{\pi}}(\mathcal{A}_{8}\mathcal{O}_{r}+\mathcal{A}_{9}\mathcal{P}_{r})
  Y(\Lambda,m_{P},r),\\
V_{V}^e &=& +\frac{\lambda_Sg_V}{6\sqrt{6}}\left(\frac{g_{BBV}}{2m_1}+\frac{g_{BBV}}{2m_3}+\frac{f_{BBV}}{m_B}\right)\nonumber\\
&&\times(2\mathcal{A}_{8}\mathcal{O}_{r}-\mathcal{A}_{9}\mathcal{P}_{r}) Y(\Lambda,m_{V},r),
\end{eqnarray}
for process ${\mathcal{B}_{6}B\to\mathcal{B}_{6}^*B}$,
\begin{eqnarray}
V_{\sigma}^f &=& g_{\sigma BB}l_S\mathcal{A}_3Y(\Lambda,m_{\sigma},r),\\
V_{P}^f &=& -\frac{g_{BBP}g_1}{2f_{\pi}m_P}(\mathcal{A}_{4}\mathcal{O}_{r}+\mathcal{A}_{5}\mathcal{P}_{r})
  Y(\Lambda,m_{P},r),\\
V_{V}^f &=& -\frac{\beta_Sg_Vg_{BBV}}{\sqrt{2}}\mathcal{A}_{3}Y(\Lambda,m_{V},r)\nonumber\\
&&-\frac{\beta_Sg_Vf_{BBV}}{8\sqrt{2}m_B}\left(\frac{1}{m_1}+\frac{1}{m_3}\right)\mathcal{A}_{3}\mathcal{O}_{r}
  Y(\Lambda,m_{V},r)\nonumber\\
  &&-\frac{\lambda_s g_V}{6\sqrt{2}}\left(\frac{g_{BBV}}{2m_1}+\frac{g_{BBV}}{2m_3}+\frac{f_{BBV}}{m_B}\right)\nonumber\\
  &&\times(2\mathcal{A}_{4}\mathcal{O}_{r}-\mathcal{A}_{5}\mathcal{P}_{r})
  Y(\Lambda,m_{V},r),
\end{eqnarray}
for process ${\mathcal{B}_{6}^*B\to\mathcal{B}_{6}^*B}$, respectively. Here, $m_1$, $m_2$, $m_3$, and $m_4$ correspond to the hadron masses from the $h_1h_2\to h_3h_4$ process, and we have defined several operators as
\begin{align}
\mathcal{O}_r &= \frac{1}{r^2}\frac{\partial}{\partial r}r^2\frac{\partial}{\partial r},\quad\quad\quad
\mathcal{P}_r = r\frac{\partial}{\partial r}\frac{1}{r}\frac{\partial}{\partial r},\\
\mathcal{A}_{1} &= \chi_{4}^{\dag}\chi_{3}^{\dag} \bm{\sigma}_1 \cdot \bm{\sigma}_2 \chi_{2}\chi_{1}, \\
\mathcal{A}_{2} &= \chi_{4}^{\dag}\chi_{3}^{\dag} S(\hat{r},\bm{\sigma_1},\bm{\sigma_2})\chi_{2}\chi_{1}, \\
\mathcal{A}_{3} &=\sum\limits_{a,b}^{m,n}C_{1/2,a;1,b}^{3/2,a+b}C_{1/2,m;1,n}^{3/2,m+n}\chi_{4a}^{\dag}\chi_{3}^{\dag}(\bm{\epsilon}_2^n\cdot\bm{\epsilon}_4^{b\dag})\chi_{2m}\chi_{1},\\
\mathcal{A}_{4} &=\sum\limits_{a,b}^{m,n}C_{{1}/{2},a;1,b}^{{3}/{2},a+b}C_{1/2,m;1,n}^{3/2,m+n}\chi_{4a}^{\dag}\chi_{3}^{\dag}i\bm{\sigma_1}\cdot(\bm{\epsilon_2^n}\times\bm{\epsilon_4^{b\dag}})\chi_{2m}\chi_{1},\\
\mathcal{A}_{5} &=\sum\limits_{a,b}^{m,n}C_{{1}/{2},a;1,b}^{{3}/{2},a+b}C_{1/2,m;1,n}^{3/2,m+n}\chi_{4a}^{\dag}\chi_{3}^{\dag}S(\hat{r},\bm{\sigma_1},i\bm{\epsilon_2^n}\times\bm{\epsilon_4^{b\dag}})\chi_{2m}\chi_{1},\\
\mathcal{A}_{6} &=\sum\limits_{m,n}C_{1/2,m;1,n}^{3/2,m+n}\chi_{4m}^{\dag}\chi_{3}^{\dag}(\bm{\sigma}_1\cdot\bm{\epsilon}_4^{n\dag})\chi_{2}\chi_{1},\\
\mathcal{A}_{7} &=\sum\limits_{m,n}C_{1/2,m;1,n}^{3/2,m+n}\chi_{4m}^{\dag}\chi_{3}^{\dag}S(\hat{r},\bm{\sigma}_1,\bm{\epsilon}_4^{n\dag})\chi_{2}\chi_{1},\\
\mathcal{A}_{8} &=\sum\limits_{m,n}C_{1/2,m;1,n}^{3/2,m+n}\chi_{4m}^{\dag}\chi_{3}^{\dag}(\bm{\sigma}_1 \cdot i\bm{\sigma}_2\times\bm{\epsilon}_4^{n\dag})\chi_{2}\chi_{1},\\
\mathcal{A}_{9} &=\sum\limits_{m,n}C_{1/2,m;1,n}^{3/2,m+n}\chi_{4m}^{\dag}\chi_{3}^{\dag}S(r,\bm{\sigma}_1,i\bm{\sigma}_2\times\bm{\epsilon}_4^{n\dag})\chi_{2}\chi_{1},
\end{align}
with $S(\hat{r},\bm{a},\bm{b}) =
3(\hat{r}\cdot\bm{a})(\hat{r}\cdot\bm{b})-\bm{a}\cdot\bm{b}$. After sandwiching these operators between the relevant spin-orbit wave functions, we can obtain the corresponding matrix elements of interaction operators, which are collected in Table \ref{operator}. 
Finally, the function $Y(\Lambda, m, r)$ reads as 
\begin{eqnarray}
Y(\Lambda, m, r) &=& \left\{\begin{array}{l}                                                    \frac{e^{-mr}-e^{-\Lambda r}}{4\pi r}-\frac{\Lambda^2-m^2}{8\pi \Lambda}e^{-\Lambda r},         \quad(\text{for}~m_E^{2}>q_i^2),\\ \nonumber\\                                                   \frac{1}{4\pi r}\left(-e^{-\Lambda r}-\frac{(\Lambda^2+m^2)r}{2\Lambda}e^{-\Lambda  r}+\text{cos}(mr)\right).\end{array}\right.        
\end{eqnarray}
\renewcommand\tabcolsep{0.2cm}
\renewcommand{\arraystretch}{1.7}
\begin{table}[!htbp]
\caption{Matrix elements for the spin-spin interactions and tensor force operators in the OBE effective potentials.}\label{operator}
{\begin{tabular}{clll}
\toprule[1pt]
\toprule[1pt]
{Spin}
       &$ \mathcal{A}_{1}$       & $\mathcal{A}_{2}$          &$ \mathcal{A}_{3}$\\\hline
$J=0$   &$\left(1\right)$
              &$\left(-4\right)$    &$\left(1\right)$\\
$J=1$    &$\left(\begin{array}{cc}-3  &0\\  0   &1 \end{array}\right)$
             &{$\left(\begin{array}{cc}0  &0\\  0   &2 \end{array}\right)$}
             &$\left(\begin{array}{cc}1  &0\\  0   &1 \end{array}\right)$  \\
$J=2$    &$\left(1\right)$
             &$\left(-\frac{2}{5}\right)$
             &$\left(\begin{array}{cc}1  &0\\  0   &1 \end{array}\right)$
                  \\
\bottomrule[1pt]
{Spin}
             &$ \mathcal{A}_{4}$
             &$ \mathcal{A}_{5}$             &$ \mathcal{A}_{6}$\\\hline
$J=0$   &$ \left(-\frac{5}{3}\right)$
                             &$ \left(\frac{2}{3}\right)$  &$ \left(2\sqrt{\frac{2}{3}}\right)$
                                 \\
$J=1$    &$\left(\begin{array}{cc}-\frac{5}{3}  &0\\  0   &1 \end{array}\right)$
             &{$\left(\begin{array}{cc}-\frac{1}{3}  &-\frac{3}{\sqrt{5}}\\ -\frac{3}{\sqrt{5}}   &-\frac{7}{5} \end{array}\right)$}
             &$\left(\begin{array}{cc}0  &2\sqrt{\frac{2}{3}}\\  0   &0 \end{array}\right)$
                                 \\
$J=2$    &$\left(\begin{array}{cc}-\frac{5}{3}  &0\\  0   &1 \end{array}\right)$
             &{$\left(\begin{array}{cc}\frac{1}{15}  &\frac{3}{5}\\ \frac{3}{5}   &\frac{7}{5} \end{array}\right)$}
             &$\left(\begin{array}{cc}2\sqrt{\frac{2}{3}}  &0 \end{array}\right)$

\\\bottomrule[1pt]
{Spin}
       &$ \mathcal{A}_{7}$
        &$ \mathcal{A}_{8}$
       &$ \mathcal{A}_{9}$\\\hline
$J=0$   &$\left(\sqrt{\frac{2}{3}}\right)$
              &$\left(-2\sqrt{\frac{2}{3}}\right)$   &$\left(-\sqrt{\frac{2}{3}}\right)$\\
$J=1$   &$\left(\begin{array}{cc}0  &-\frac{1}{\sqrt{6}}\\  -2\sqrt{\frac{3}{5}}   &-3\sqrt{\frac{3}{10}} \end{array}\right)$
             &$\left(\begin{array}{cc}0  &-2\sqrt{\frac{2}{3}}\\  0   &0 \end{array}\right)$
             &$\left(\begin{array}{cc}0  &\frac{1}{\sqrt{6}}\\  2\sqrt{\frac{3}{5}}   &3\sqrt{\frac{3}{10}} \end{array}\right)$  \\
$J=2$   &$\left(\begin{array}{cc}\frac{1}{5\sqrt{6}}  &\frac{3\sqrt{6}}{10}\\   \end{array}\right)$
            &$\left(\begin{array}{cc}-2\sqrt{\frac{2}{3}}  &0\\   \end{array}\right)$
             &$\left(\begin{array}{cc}-\frac{1}{5\sqrt{6}}  &-\frac{3\sqrt{6}}{10}\\   \end{array}\right)$
                             \\\bottomrule[1pt]\bottomrule[1pt]
\end{tabular}}
\end{table}

\section{Numerical results}\label{sec3}

After setting up the systems, we proceed to search for possible molecular and resonant di-baryons. We solve the coupled channel Schr\"{o}dinger equations to identify reasonable bound states, the results are collected in Sec.~\ref{bound} including binding energies $E$ that range from several to several tens of MeV and a root-mean-square (RMS) radius of approximately 1.00 fm or larger. For the resonances, we analyze the phase shifts for all discussed channels, as detailed in Sec.~\ref{resonant}.

It is worthy to note that in our work, the cutoff $\Lambda$ is the only free parameter, which is varied in the region of $0.80\leq\Lambda\leq2.00$ GeV. Usually, a reasonable cutoff  around 1.00 GeV is selected, which is based on experience with nucleon-nucleon interactions \cite{Tornqvist:1993ng,Tornqvist:1993vu}.

\subsection{Bound states}\label{bound}

\paragraph{\bf{The $\Xi_cN$ systems.}}
The quantum number configurations for the $\Xi_cN$ systems encompass $I(J^P)=0(0^-)$, $0(1^-)$, $0(2^-)$, $1(0^-)$, $1(1^-)$, and $1(2^-)$. As indicated in Table \ref{obe}, the OBE effective potentials for the single $\Xi_cN$ systems via the $P-$wave interactions are identical to those in $S-$wave interactions \cite{Huo:2024eew}, but the presence of a repulsive centrifugal force renders the search for bound state solutions in the single $\Xi_cN$ case particularly challenging.

\renewcommand{\arraystretch}{2.0}
\begin{table*}[!hbtp]
\renewcommand\tabcolsep{0.1cm}
\renewcommand{\arraystretch}{1.6}
\caption{The bound state solutions (the binding energy $E$, the root-mean-square radius $r_{RMS}$, and the probabilities $p_i$ for all the discussed channels) for the $\Xi_cN$ systems with $I(J^P)=0(0^-)$, $0(1^-)$, $0(2^-)$, $1(0^-)$, $1(1^-)$, and $1(2^-)$ after considered the coupled channel effects. Here, the unites for the cutoff $\Lambda$, the binding energy $E$, and the root-mean-square radius $r_{RMS}$ are GeV, MeV, and fm, respectively. }\label{XicN}
\begin{tabular}{ccccccccccccc}
\toprule[1.5pt]
\toprule[1.5pt]
$I(J^P)$  &$\Lambda$    &$E$    &$r_{RMS}$     &\multicolumn{1}{c}{$\Xi_cN(^{1}P/^{3}P)$}    &$\Lambda_c\Sigma(^{1}P/^{3}P)$  &$\Xi_c^{\prime}N(^{1}P/^{3}P)$     &$\Sigma_c\Lambda(^{1}P/^{3}P)$ &$\Xi_c^{*}N(^{3}P/^{5}P)$      &$\Sigma_c^{*}\Lambda(^{3}P/^{5}P)$ &$\Sigma_c\Sigma(^{1}P/^{3}P)$        &$\Sigma_c^{*}\Sigma(^{3}P/^{5}P)$  \\  \hline
$0(0^-)$             & 1.108    &$-$1.70        &1.23           &\multicolumn{1}{c}{--/\bf{53.90}}   &--/--    &--/8.08     &--/--    &{\bf{37.95}}/--     &--/--     &--/0.07   &0.01/--          \\
                      & 1.110   &$-$5.04        &0.93          &\multicolumn{1}{c}{--/{\bf{50.56}}}   &--/--    &--/8.36    &--/--    &{\bf{41.00}}/--    &--/--      &--/0.07   &0.01/--          \\
                       & 1.112   &$-$8.63        &0.82          &\multicolumn{1}{c}{--/{\bf{48.24}}} &--/--  &--/8.48    &--/--    &{\bf{43.20}}/--    &--/--      &--/0.07   &0.01/--          \\ \hline

$0(1^-)$  &1.074    &$-$0.14        &2.05           &\multicolumn{1}{c}{{\bf{31.59}/{\bf{20.52}}}}  &--/-- &{\bf{31.55}}/3.88     &--/--   &{\bf{12.13}}/0.02    &--/--      &0.13/0.13   &0.05/$~\sim$0.00          \\
                      &1.076   &$-$3.22        &0.97          &\multicolumn{1}{c}{{\bf{29.14}}/{\bf{18.84}}} &--/--  &{\bf{34.20}}/4.24   &--/--     &{\bf{13.19}}/0.02   &--/--       &0.15/0.15   &0.06/$~\sim$0.00          \\
                       &1.078   &$-$6.52        &0.83          &\multicolumn{1}{c}{{\bf{27.93}}/{\bf{18.05}}} &--/--  &{\bf{35.44}}/4.43    &--/--    &{\bf{13.74}}/0.02    &--/--      &0.16/0.17   &0.06/$~\sim$0.00          \\ \hline
$0(2^-)$  &1.108    &$-$0.05       &2.53           &\multicolumn{1}{c}{--/{\bf{52.40}}}  &--/-- &--/6.50    &--/--    &{\bf{41.00}}/0.09    &--/--      &--/0.01   &0.01/0.01          \\
                       &1.110   &$-$3.57        &0.92          &\multicolumn{1}{c}{--/\bf{47.62}} &--/--  &--/7.10   &--/--     &{\bf{45.18}}/0.08    &--/--      &--/0.01   &0.01/0.01          \\
                       & 1.112   &$-$7.35        &0.78          &--/\bf{45.45}  &--/-- &--/7.34   &--/--     &{\bf{47.11}}/0.07  &--/--        &--/0.01   &0.01/0.01          \\ \hline
 $1(0^-)$                &1.103    &$-$1.84        &0.55           &\multicolumn{1}{c}{--/5.66}   &--/\bf{34.88}        &--/0.09          &--/3.76   &2.45/-- &{\bf{20.31}}/--  &--/4.40   &\bf{28.45}/--   \\
                       &1.104   &$-$4.71        &0.52          &\multicolumn{1}{c}{--/5.38}   &--/{\bf{34.74}}        &--/0.09          &--/3.76   &2.48/--   &{\bf{20.48}}/--     &--/4.40  &{\bf{28.67}}/--\\
                        &1.105   &$-$7.61        &0.50          &\multicolumn{1}{c}{--/5.20}   &--/{\bf{34.57}}        &--/0.09          &--/3.75   &2.50/--   &{\bf{20.63}}/--   &--/4.40    &{\bf{28.86}}/--       \\ \hline
$1(1^-)$             &1.092    &$-$3.83        &0.51           &\multicolumn{1}{c}{4.21/0.01}   &{\bf{29.70}}/$~\sim$0.00        &2.68/$~\sim$0.00          &{\bf{25.58}}/$~\sim$0.00   &$~\sim$0.00/0.03 &$~\sim$0.00/0.96  &{\bf{36.72}}/$~\sim$0.00   &$~\sim$0.00/0.11   \\
                      & 1.093   &$-$6.68        &0.50          &\multicolumn{1}{c}{4.09/$~\sim$0.01}   &{\bf{29.57}}/$~\sim$0.00        &2.70/$~\sim$0.00          &{\bf{25.68}}/$~\sim$0.00   &$~\sim$0.00/0.03   &$~\sim$0.00/0.97     &{\bf{36.85}}/$~\sim$0.00  &$~\sim$0.00/0.10\\
                       &1.094   &$-$9.56        &0.49          &\multicolumn{1}{c}{4.00/0.01}   &{\bf{29.44}}/$~\sim$0.00        &2.71/$~\sim$0.00          &{\bf{25.76}}/$~\sim$0.00   &$~\sim$0.00/0.03   &$~\sim$0.00/0.98   &{\bf{36.96}}/$~\sim$0.00    &$~\sim$0.00/0.10       \\ \hline
 $1(2^-)$                      &1.103    &$-$1.73        &0.54           &\multicolumn{1}{c}{--/5.31}   &--/{\bf{33.80}}        &--/0.28          &--/3.15   &2.31/0.02 &{\bf{21.29}}/0.01  &--/4.23   &{\bf{29.57}}/0.03  \\
                      & 1.104   &$-$4.64        &0.51          &\multicolumn{1}{c}{--/5.07}   &--/{\bf{33.69}}        &--/0.28          &--/3.16   &2.33/0.02   &{\bf{21.42}}/0.01     &--/4.25  &{\bf{29.74}}/0.03\\
                       & 1.105   &$-$7.58        &0.49          &\multicolumn{1}{c}{--/4.93}   &--/{\bf{33.55}}        &--/0.29          &--/3.17   &2.35/0.02   &{\bf{21.53}}/0.01   &--/4.26    &{\bf{29.88}}/0.03       \\
                       \bottomrule[1.5pt]
\bottomrule[1.5pt]
\end{tabular}
\end{table*}

After a single channel investigation, we further consider the coupled channel effects including $\Xi_cN$, $\Lambda_c\Sigma$, $\Xi_c^{\prime}N$, $\Sigma_c\Lambda$, $\Xi_c^*N$, $\Sigma_c^*\Lambda$, $\Sigma_c\Sigma$, and $\Sigma_c^*\Sigma$. In Table \ref{XicN}, we collect the corresponding bound state solutions in the cutoff region $\Lambda<2.00$ GeV. As we can see, bound state solutions now exist for all coupled channel systems when the cutoff is taken around 1.00 GeV. However, the dominant channels are not the lowest channel $\Xi_cN$, leading to the small size of these bound states. Additionally, the obtained binding energies are highly sensitive to the cutoff value. Thus, we cannot recommend these coupled channel bound states as primary molecular candidates.

Nevertheless, several coupled interactions demonstrate sufficiently strong attractive forces: the $\Xi_cN/\Xi_c^*N$ with $0(0^-)$, the $\Xi_cN/\Xi_c^{\prime}N/\Xi_c^*N$ with $0(1^-)$, the $\Xi_cN/\Xi_c^*N$ with $0(2^-)$, the $\Lambda_c\Sigma/\Sigma_c^*\Lambda/\Sigma_c^*\Sigma$ with $1(0^-)$, the $\Lambda_c\Sigma/\Sigma_c\Lambda/\Sigma_c\Sigma$ with $1(1^-)$, and the $\Lambda_c\Sigma/\Sigma_c^*\Lambda/\Sigma_c^*\Sigma$ with $1(2^-)$. These channels exhibit large probabilities in their respective coupled systems. Thus, there is potential to search for charm-strange resonant dibaryons, and we will perform a phase shift analysis on these coupled channels in Sec. \ref{resonant}.

\paragraph{\bf{The $\Lambda_c\Sigma$ systems.}}
For the $\Lambda_c\Sigma$ systems, the discussed quantum number configurations include $I(J^P)=1(0^-)$, $1(1^-)$, and $1(2^-)$. As shown in Table \ref{obe}, only the $\sigma$ and $\omega$ exchanges interactions are permitted, and these interactions are independent of the spin-parities. This finding aligns with the conclusions presented in Ref.~\cite{Huo:2024eew}. In the cutoff range of $0.80\leq\Lambda\leq2.00$ GeV, we find no loosely bound states for the single $\Lambda_c\Sigma$ systems, suggesting the OBE interactions lack the necessary attractive strength to form such systems.

\renewcommand{\arraystretch}{2.0}
\begin{table*}[!hbtp]
\renewcommand\tabcolsep{0.18cm}
\renewcommand{\arraystretch}{1.6}
\caption{The bound state solutions (the binding energy $E$, the root-mean-square radius $r_{RMS}$, and the probabilities $p_i$ for all the discussed channels) for the $\Lambda_c\Sigma$ systems with $I(J^P)=1(0^-)$, $1(1^-)$, and $1(2^-)$. Here, the unites for the cutoff $\Lambda$, the binding energy $E$, and the root-mean-square radius $r_{RMS}$ are GeV, MeV, and fm, respectively. }\label{LambdacSigma}
\begin{tabular}{ccccccccccccc}
\toprule[1.5pt]
\toprule[1.5pt]
     $I(J^P)$                  &$\Lambda$    &$E$    &$r_{RMS}$   &$\Lambda_c\Sigma(^{1}P/^{3}P)$  &$\Xi_c^{\prime}N(^{1}P/^{3}P)$     &$\Sigma_c\Lambda(^{1}P/^{3}P)$ &$\Xi_c^{*}N(^{3}P/^{5}P)$      &$\Sigma_c^{*}\Lambda(^{3}P/^{5}P)$ &$\Sigma_c\Sigma(^{1}P/^{3}P)$        &$\Sigma_c^{*}\Sigma(^{3}P/^{5}P)$  \\  \hline
    $1(0^-)$                   &1.098  &$-$1.41        &0.51           &--/{\bf{35.60}}        &--/0.05          &--/3.28   &1.57/--      &{\bf{21.77}}/--   &--/5.20   &{\bf{32.53}}/--    \\
                        &1.100   &$-$6.18        &0.50           &--/{\bf{35.25}}        &--/0.04          &--/3.25   &1.59/--      &{\bf{21.90}}/--   &--/5.17   &{\bf{32.78}}/--    \\
                        &1.102   &$-$11.06        &0.50       &--/{\bf{34.93}}        &--/0.04          &--/3.23   &1.61/--   &{\bf{22.03}}/--    &--/5.14      &{\bf{33.02}}/--       \\ \hline
     $1(1^-)$     &1.072   &$-$6.23        &0.79          &{\bf{43.76}}/0.11        &1.20/$~\sim$0.00          &{\bf{21.05}}/$~\sim$0.00   &$~\sim$0.00/0.04      &0.01/0.84   &{\bf{32.70}}/$~\sim$0.00   &0.03/0.25    \\
                        &1.074   &$-$9.83        &0.71          &{\bf{42.08}}/0.08        &1.25/$~\sim$0.00          &{\bf{21.66}}/$~\sim$0.00   &$~\sim$0.00/0.04  &0.01/0.88   &{\bf{33.73}}/$~\sim$0.00  &0.03/0.24        \\
                        &1.076   &$-$13.57        &0.67          &{\bf{40.75}}/0.06       &1.28/0.00          &{\bf{22.13}}/$~\sim$0.00   &$~\sim$0.00/0.04   &$~\sim$0.00/0.92    &{\bf{34.55}}/$~\sim$0.00      &0.03/0.22       \\ \hline
     $1(2^-)$    &1.084   &$-$1.77        &1.06       &--/{\bf{52.73}}        &--/0.10          &--/2.46   &0.87/0.03      &{\bf{15.80}}/0.01   &--/3.72   &{\bf{24.28}}/0.02    \\
                       &1.086   &$-$5.22        &0.83          &--/{\bf{50.05}}        &--/0.10          &--/2.58   &0.92/0.03  &{\bf{16.69}}/0.01   &--/3.91  &{\bf{25.68}}/0.02        \\
                        &1.088   &$-$8.86        &0.73         &--/{\bf{48.23}}        &--/0.11         &--/2.66   &0.97/0.04   &{\bf{17.30}}/0.01    &--/4.04      &{\bf{26.64}}/0.02       \\

                       \bottomrule[1.5pt]
\bottomrule[1.5pt]
\end{tabular}
\end{table*}

In Table \ref{LambdacSigma}, we present the numerical results for the $\Lambda_c\Sigma/\Xi_c^{\prime}N/\Sigma_c\Lambda/\Xi_c^*N/\Sigma_c^*\Lambda/\Sigma_c\Sigma/\Sigma_c^*\Sigma$ coupled interactions with quantum numbers $I(J^P)=1(0^-)$, $1(1^-)$, and $1(2^-)$. These results closely resemble the bound state solutions reported in Table \ref{XicN}. Notably, the cutoffs employed here are slightly smaller than those in Table \ref{XicN}, and our findings indicate that the $\Xi_cN$ channels play a positive, albeit not significantly important role in the $\Xi_cN\Lambda_c\Sigma/\Xi_c^{\prime}N/\Sigma_c\Lambda/\Xi_c^*N/\Sigma_c^*\Lambda/\Sigma_c\Sigma/\Sigma_c^*\Sigma$ coupled interactions with the specified quantum numbers.

Furthermore, although bound state solutions for the $\Lambda_c\Sigma$ systems with $I(J^P)=1(0^-)$, $1(1^-)$, and $1(2^-)$ can be obtained when the cutoff is set around 1.00 GeV, these coupled bound states are not ideal molecular candidates. This conclusion is supported by the sensitivity of the binding energies to the cutoff $\Lambda$, and the smaller RMS radii, which are less than the reasonable size expected for hadronic molecules.

\paragraph{\bf{The $\Xi_c^{\prime}N$ systems.}}
We still first perform the single channel analysis on the $\Xi_c^{\prime}N$ systems. After solving the coupled channel Schr\"{o}dinger equations, we can obtain loosely bound state solutions for the $\Xi_c^{\prime}N$ state with $0(1^-)$ at a cutoff of approximately 1.00 GeV. In this case, the dominant channel is $\Xi_c^{\prime}N$. In Table \ref{XicpN}, we collect the corresponding numerical results. After incorporating the coupled-channel effects, we find the coupled $\Xi_c^{\prime}N/\Sigma_c\Lambda/\Xi_c^{*}N/\Sigma_c^*\Lambda/\Sigma_c\Sigma/\Sigma_c^*\Sigma$ state with $0(1^-)$ becomes more tightly bound than the single $\Xi_c^{\prime}N$, with the $\Xi_cN$ channel now emerging as the dominant interaction. This indicates that the coupled channel effects play a significant role in enhancing binding. In conclusion, due to the reasonable RMS radius and cutoff value, this bound state can be regarded as a promising molecular candidate.

For the remaining $\Xi_c^{\prime}N$ systems with $I(J^P) = 0(0^-)$, $0(2^-)$, $1(0^-)$, $1(1^-)$, and $1(2^-)$, our results indicate that the OBE effective potentials can not generate sufficiently strong attractive interactions to obtain loosely bound states at the cutoff value of approximately 1.00 GeV. After accounting for coupled-channel effects, loosely bound states do emerge, but their RMS radii remain around 0.50 fm, which is significantly smaller than the typical size of hadronic molecules. Hence, despite the stronger interactions introduced by the coupled channel effects, these five $\Xi_c^{\prime}N$ bound states cannot be considered viable molecular candidates. However, with sufficiently strong interactions, resonance states may arise from the coupled interactions between $\Xi_c^*N/\Sigma_c^*\Sigma$ ($0(0^-, 2^-)$), $\Sigma_c^*\Lambda/\Sigma_c^*\Sigma$ ($1(0^-, 2^-)$), and $\Sigma_c\Lambda/\Sigma_c\Sigma$ ($1(1^-)$) channels, which will be discussed in Section \ref{resonant}.

\renewcommand{\arraystretch}{2.0}
\begin{table*}[!hbtp]
\renewcommand\tabcolsep{0.17cm}
\renewcommand{\arraystretch}{1.5}
\caption{The bound state solutions (the binding energy $E$, the root-mean-square radius $r_{RMS}$, and the probabilities $p_i$ for all the discussed channels) for the $\Xi_c^{\prime}N$ systems with $I(J^P)=0(0^-)$, $0(1^-)$, $0(2^-)$, $1(0^-)$, $1(1^-)$, and $1(2^-)$ after considered the coupled channel effects. Here, the unites for the cutoff $\Lambda$, the binding energy $E$, and the root-mean-square radius $r_{RMS}$ are GeV, MeV, and fm, respectively. }\label{XicpN}
\begin{tabular}{ccccccccccccc}
\toprule[1.5pt]
\toprule[1.5pt]
Single channel &$I(J^P)$  &$\Lambda$    &$E$    &$r_{RMS}$   &$\Xi_c^{\prime}N(^{1}P/^{3}P)$
       &\multicolumn{1}{|c}{$I(J^P)$}  &$\Lambda$    &$E$    &$r_{RMS}$
 &$\Xi_c^{\prime}N(^{1}P/^{3}P)$  \\ \cline{2-11}
 &$0(0^-)$  &--/--   &--/--&--/--&--/--      &\multicolumn{1}{|c}{$0(2^-)$} &--/--   &--/--&--/--&--/--     \\
&$0(1^-)$   &1.217    &$-$0.27        &1.93          &100/0.00
    &\multicolumn{1}{|c}{$1(0^-)$} &--/--   &--/--&--/--&--/-- \\
                      && 1.222    &$-$4.96        &0.99         &100/0.00
   &\multicolumn{1}{|c}{$1(1^-)$} &--/--   &--/--&--/--&--/-- \\
                       &&1.227    &$-$10.20        &0.84         &100/0.00
  &\multicolumn{1}{|c}{$1(2^-)$} &--/--   &--/--&--/--&--/-- \\ \midrule[1.5pt]
Couple channel &$I(J^P)$  &$\Lambda$    &$E$    &$r_{RMS}$     &$\Xi_c^{\prime}N(^{1}P/^{3}P)$     &$\Sigma_c\Lambda(^{1}P/^{3}P_J)$ &$\Xi_c^{*}N(^{3}P/^{5}P)$      &$\Sigma_c^{*}\Lambda(^{3}P/^{5}P)$ &$\Sigma_c\Sigma(^{1}P/^{3}P)$        &$\Sigma_c^{*}\Sigma(^{3}P/^{5}P)$  \\  \cline{2-11}
&$0(0^-)$             &1.226   &$-$1.18        &0.51           &--/2.63     &--/--    &{\bf{41.95}}/--          &--/--  &--/1.99   &{\bf{53.42}}/--          \\
                        &&1.229   &$-$6.61        &0.51         &--/2.68     &--/--    &{\bf{41.46}}/--   &--/--        &--/2.04   &{\bf{53.81}}/--     \\
                        &&1.232   &$-$12.19        &0.50        &--/2.74    &--/--     &{\bf{41.05}}/--     &--/--      &--/2.09   &{\bf{54.12}}/--          \\ \cline{2-11}

&$0(1^-)$      &1.160   &$-$0.90        &2.28            &{\bf{49.03/45.52}}    &--/--      &1.34/2.01          &--/--  &1.78/0.13   &0.13/0.02          \\
                        &&1.170   &$-$4.38        &1.48            &{\bf{57.40/36.82}}    &--/--      &1.73/2.17           &--/--   &1.72/0.02   &0.08/0.01          \\
                        &&1.180   &$-$9.07        &1.19              &{\bf{64.07/30.06}}        &--/--  &2.00/2.13            &--/--  &1.65/0.00   &0.04/0.01  \\ \cline{2-11}
&$0(2^-)$     &1.230   &$-$3.14        &0.58          &--/7.94   &--/--       &{\bf{41.14}}/1.09    &--/--        &--/5.75   &{\bf{42.51}}/1.55          \\
                       & &1.232   &$-$6.71        &0.55          &--/7.49   &--/--       &{\bf{41.09}}/1.01        &--/--    &--/5.81   &{\bf{43.06}}/1.53          \\
                      & & 1.234   &$-$10.35        &0.53          &--/7.19   &--/--       &{\bf{41.00}}/0.94         &--/--   &--/5.86   &{\bf{43.51}}/1.51         \\ \cline{2-11}
& $1(0^-)$             &1.282     &$-$1.80        &0.45                &--/0.17          &--/4.97   &5.11/--    &{\bf{37.83}}/--   &--/1.76     &{\bf{50.17}}/--     \\
                       & &1.283     &$-$4.01        &0.45              &--/0.17          &--/4.98   &5.11/--     &{\bf{37.82}}/--   &--/1.77    &{\bf{50.15}}/--    \\
                        &&1.284   &$-$6.23        &0.45                &--/0.17          &--/4.98   &5.11/--      &{\bf{37.82}}/--     &--/1.79    &{\bf{50.13}}/--   \\ \cline{2-11}
&$1(1^-)$             &1.220     &$-$0.12        &0.70               &5.24/0.04          &{\bf{38.34}}/0.01   &0.05/0.17    &0.12/5.38   &{\bf{50.19}}/0.03     &0.20/0.19     \\
                       & &1.223     &$-$5.70        &0.53            &4.74/0.03          &{\bf{38.43}}/0.01   &0.05/0.17     &0.12/5.42   &{\bf{51.61}}/0.03    &0.20/0.18    \\
                      & & 1.226   &$-$11.45        &0.51                 &4.59/0.02          &{\bf{38.39}}/0.01  &0.05/0.16      &0.11/5.44     &{\bf{50.81}}/0.03     &0.18/0.18    \\ \cline{2-11}
& $1(2^-)$        &1.284     &$-$0.91        &0.46           &--/0.63          &--/5.97   &4.60/0.09    &{\bf{37.07}}/0.17   &--/5.97     &{\bf{44.30}}/1.19     \\
                        &&1.286     &$-$5.42        &0.45         &--/0.60          &--/5.94   &4.62/0.08     &{\bf{37.13}}/0.15   &--/5.96    &{\bf{44.34}}/1.16    \\
                      & &1.288   &$-$10.00        &0.44              &--/0.59          &--/5.91   &4.65/0.07      &{\bf{37.19}}/0.14     &--/5.95     &{\bf{44.37}}/1.13    \\\bottomrule[1.5pt]
\bottomrule[1.5pt]
\end{tabular}
\end{table*}

\paragraph{\bf{The $\Sigma_c\Lambda$ systems.}}
For the single $\Sigma_c\Lambda$ systems with $1(0^-, 1^-, 2^-)$, we fail to obtain loosely bound state solutions for cutoff values $\Lambda\leq2.00$ GeV, as shown in Table \ref{SigmacLambda}. We then investigate the coupled channel interactions involving $\Sigma_c\Lambda/\Xi_c^*N/\Sigma_c^*\Lambda/\Sigma_c\Sigma/\Sigma_c^*\Sigma$ for the same quantum numbers, and the results shows again the $\Sigma_c^*\Lambda/\Sigma_c^*\Sigma$ coupled interactions with $1(0^-, 2^-)$, as well as the $\Sigma_c\Lambda/\Sigma_c^*\Lambda$ coupled interactions with $1(1^-)$, are sufficiently strong to generate bound states. However, none of these bound states can be considered molecular candidates primarily composed of the $\Sigma_c\Lambda$ system.

\renewcommand{\arraystretch}{2.0}
\begin{table*}[!hbtp]
\renewcommand\tabcolsep{0.3cm}
\renewcommand{\arraystretch}{1.6}
\caption{The bound state solutions (the binding energy $E$, the root-mean-square radius $r_{RMS}$, and the probabilities $p_i$ for all the discussed channels) for the $\Sigma_c\Lambda$ systems with $I(J^P)=1(0^-)$, $1(1^-)$, and $1(2^-)$ after considered the coupled channel effects. Here, the unites for the cutoff $\Lambda$, the binding energy $E$, and the root-mean-square radius $r_{RMS}$ are GeV, MeV, and fm, respectively. }\label{SigmacLambda}
\begin{tabular}{ccccccccccccc}
\toprule[1.5pt]
\toprule[1.5pt]
     $I(J^P)$                  &$\Lambda$    &$E$    &$r_{RMS}$    &$\Sigma_c\Lambda(^{1}P/^{3}P)$ &$\Xi_c^{*}N(^{3}P/^{5}P)$      &$\Sigma_c^{*}\Lambda(^{3}P/^{5}P)$ &$\Sigma_c\Sigma(^{1}P/^{3}P)$        &$\Sigma_c^{*}\Sigma(^{3}P/^{5}P)$  \\  \hline
    $1(0^-)$        &1.256     &$-$0.63        &0.56              &--/5.53   &5.59/--     &{\bf{37.62}}/--   &--/1.40    &{\bf{49.86}}/--    \\
                   &1.259     &$-$5.86        &0.52       &--/5.35   &5.55/--    &{\bf{37.62}}/--   &--/1.44      &{\bf{50.03}}/--     \\
                   & 1.262     &$-$11.25        &0.51    &--/5.30   &5.53/--   &{\bf{37.58}}/--     &--/1.49     &{\bf{50.10}}/--      \\ \hline
     $1(1^-)$     &1.220     &$-$0.93        &1.03         &{\bf{41.96}}/0.05   &0.11/0.20     &0.30/5.60   &{\bf{50.84}}/0.06    &0.49/0.33    \\
      & 1.225     &$-$6.75        &0.71          &{\bf{39.08}}/0.04  &0.11/0.18   &0.31/5.93     &{\bf{53.49}}/0.06     &0.49/0.32  \\
                       & 1.230     &$-$13.00        &0.63              &{\bf{37.72}}/0.04   &0.10/0.15    &0.29/6.10   &{\bf{54.75}}/0.05      &0.46/0.29     \\
                       \hline
     $1(2^-)$    &1.260     &$-$1.11        &0.62               &--/8.27   &4.95/0.22     &{\bf{35.39}}/0.43   &--/6.34    &{\bf{42.58}}/1.78    \\
                        &1.263     &$-$6.45        &0.54          &--/7.56   &5.00/0.19    &{\bf{35.72}}/0.38   &--/6.39      &{\bf{43.06}}/1.71     \\
                       & 1.266     &$-$11.96        &0.52             &--/7.22   &5.02/0.17   &{\bf{35.88}}/0.34     &--/6.39     &{\bf{43.34}}/1.64      \\
    \bottomrule[1.5pt]
\bottomrule[1.5pt]
\end{tabular}
\end{table*}

\paragraph{\bf{The $\Xi_c^{*}N$ systems.}}
Table \ref{XicSN} presents the bound state solutions for the $\Xi_c^*N$ states with $0(0^-)$, $0(1^-)$, $0(2^-)$, and $1(2^-)$, both for the single-channel and coupled-channel cases. We observe that, for a cutoff value $\Lambda \leq 2.00$ GeV, bound state solutions are obtained for the single $\Xi_c^*N$ states with $0(0^-)$, $0(1^-)$, and $0(2^-)$. The corresponding RMS radii are around 1.00 fm or larger, with binding energies varying from several MeV to several tens of MeV. Notably, the OBE effective potentials are significantly stronger for the isoscalar $\Xi_c^*N$ states compared to the isovector $2^-$ state, primarily due to the smaller cutoff values associated with the isoscalar channels. Additionally, the results indicate that, for the isoscalar $\Xi_c^*N$ states, the low-spin components dominate, whereas, for the isovector $\Xi_c^*N$ state with $2^-$, the dominant channel is the $\Xi_c^*N|{}^5P_2\rangle$ configuration.

When the coupled channel effects are considered, the bound state solutions for the $\Xi_c^*N$ states with $0(0^-)$, $0(1^-)$, and $0(2^-)$ exhibit minimal change compared to the single channel case, suggesting that the coupled channel effects have a negligible influence on these isoscalar states.

For the isovector $\Xi_c^*N$ systems, the loosely bound state solutions emerge at a cutoff of around 1.00 GeV. However, the dominant channels are not the $\Xi_c^*N$ systems systems themselves, but rather higher-mass systems such as $\Sigma_c^*\Sigma$, $\Sigma_c^*\Lambda$ and $\Sigma_c\Sigma$. Consequently, the RMS radii for these states are approximately 0.50 fm or smaller, which is inconsistent with the typical size of hadronic molecular states. Therefore, we conclude that the $\Xi_c^*N/\Sigma_c^*\Lambda/\Sigma_c\Sigma/\Sigma_c^*\Sigma$ coupled states with $1(0^-)$, $1(1^-)$ and $1(2^-)$ cannot be good molecular candidates.

In summary, based on the reasonable cutoff values and the bound state solutions obtained, we predict the existence of single $\Xi_c^*N$ molecular states with $0(0^-)$, $0(1^-)$, $0(2^-)$, and $1(2^-)$.

\renewcommand{\arraystretch}{2.0}
\begin{table*}[!hbtp]
\renewcommand\tabcolsep{0.2cm}
\renewcommand{\arraystretch}{1.5}
\caption{The bound state solutions (the binding energy $E$, the root-mean-square radius $r_{RMS}$, and the probabilities $p_i$ for all the discussed channels) for the $\Xi_c^*N$ systems with $I(J^P)=0(0^-)$, $0(1^-)$, $0(2^-)$, $1(0^-)$, $1(1^-)$, and $1(2^-)$. Here, the unites for the cutoff $\Lambda$, the binding energy $E$, and the root-mean-square radius $r_{RMS}$ are GeV, MeV, and fm, respectively. }\label{XicSN}
\begin{tabular}{ccccccccccc}
\toprule[1.5pt]
\toprule[1.5pt]
Single channel  &$I(J^P)$  &$\Lambda$    &$E$    &$r_{RMS}$      &$\Xi_c^{*}N(^{3}P/^{5}P)$
                &\multicolumn{1}{|c}{$I(J^P)$}  &$\Lambda$    &$E$    &$r_{RMS}$      &$\Xi_c^{*}N(^{3}P/^{5}P)$ \\\cline{2-11}
               &$0(0^-)$  &1.285    &$-$1.37        &1.59          &100/--
               &\multicolumn{1}{|c}{$0(1^-)$}  &1.340    &$-$2.22        &1.31   &91.29/8.70 \\
                 &&1.290    &$-$3.90        &1.22         &100/--
                   &\multicolumn{1}{|c}{}&1.345    &$-$6.50        &1.00   &93.08/6.91  \\
                & &1.295    &$-$6.71        &1.06         &100/--
                 &\multicolumn{1}{|c}{}& 1.350    &$-$11.30        &0.86   &94.25/5.74  \\\cline{2-11}
      &$0(2^-)$  &1.320    &$-$0.21        &2.99          &74.75/25.24
       &\multicolumn{1}{|c}{$1(0^-)$}   &--/--   &--/--&--/--&--/--            \\
                 &&1.327    &$-$4.27        &1.27         &86.21/13.78
                  &\multicolumn{1}{|c}{$1(1^-)$}   &--/--   &--/--&--/--&--/--        \\
                & &1.334    &$-$9.47        &0.98         &91.19/8.80
                &\multicolumn{1}{|c}{$1(2^-)$}   &--/--   &--/--&--/--&--/--      \\\midrule[1.5pt]

Coupled channel  &$I(J^P)$  &$\Lambda$    &$E$    &$r_{RMS}$      &$\Xi_c^{*}N(^{3}P/^{5}P)$      &$\Sigma_c^{*}\Lambda(^{3}P/^{5}P)$ &$\Sigma_c\Sigma(^{1}P/^{3}P)$        &$\Sigma_c^{*}\Sigma(^{3}P/^{5}P)$  \\  \cline{2-11}
&$0(0^-)$           &1.280     &$-$0.06        &3.35                  &{\bf{94.47}}/-- &--/--         &--/$~\sim$0.00   &5.52/--         \\
                      & & 1.290     &$-$5.11        &1.13                 &{\bf{93.03}}/--  &--/--        &--/$\sim$0.00   &6.96/--         \\
                      & & 1.300   &$-$11.21        &0.93                 &{\bf{92.15}}/--   &--/--       &--/$\sim$0.00   &7.83/--          \\ \cline{2-11}

&$0(1^-)$      &1.339     &$-$1.73        &1.41                &{\bf{87.64}}/8.94   &--/--       &0.29/0.09   &2.75/0.26         \\
                       & &1.342     &$-$4.08        &1.13                 &{\bf{88.60}}/7.68  &--/--        &0.32/0.10   &2.99/0.28         \\
                       && 1.345   &$-$6.62        &1.00                  &{\bf{89.24}}/6.79 &--/--         &0.34/0.12   &3.19/0.29          \\  \cline{2-11}
&$0(2^-)$      &1.323     &$-$0.59        &2.24                  &{\bf{71.52}}/22.25   &--/--       &--/0.07   &5.77/0.36         \\
                       & &1.329     &$-$4.01        &1.28                 &{\bf{78.44}}/13.63  &--/--        &--/0.09   &7.37/0.44         \\
                       & &1.335   &$-$8.18        &1.02                  &{\bf{81.63}}/9.32   &--/--       &--/0.10   &8.45/0.49          \\ \cline{2-11}
& $1(0^-)$            & 1.276    &$-$1.25        &0.60              &6.28/--     &{\bf{43.15}}/--   &--/4.83    &{\bf{45.74}}/--    \\
                      &  &1.279     &$-$5.72        &0.55             &5.90/--    &{\bf{43.20}}/--   &--/4.95      &{\bf{45.94}}/--     \\
                     &  & 1.282     &$-$10.33        &0.53             &5.73/--   &{\bf{43.18}}/--     &--/5.06     &{\bf{46.02}}/--      \\ \cline{2-11}
&$1(1^-)$             &1.604     &$-$2.21        &0.53              &0.17/0.75    &0.87/{\bf{29.66}}   &{\bf{54.39}}/4.13      &0.06/9.96     \\
                       & &1.607     &$-$5.84        &0.51             &0.15/0.68   &0.88/{\bf{29.91}}     &{\bf{54.02}}/4.10     &0.07/10.18      \\
 &  &1.610     &$-$9.56        &0.50              &0.13/0.63     &0.88/{\bf{30.15}}   &{\bf{53.64}}/4.06    &0.07/10.40    \\
                      \cline{2-11}
& $1(2^-)$     &1.277     &$-$1.24        &0.59              &5.12/0.23    &{\bf{39.99}}/0.52   &14.43/--      &{\bf{37.56}}/2.16     \\

& & 1.280     &$-$6.04        &0.54              &4.87/0.17     &{\bf{40.13}}/0.45   &14.44/--    &{\bf{37.83}}/2.07    \\
& &1.283     &$-$11.01        &0.53             &4.79/0.14  &{\bf{40.22}}/0.40     &14.43/--     &{\bf{38.03}}/2.00      \\                       \bottomrule[1.5pt]
\bottomrule[1.5pt]
\end{tabular}
\end{table*}

\paragraph{\bf{The $\Sigma_c^*\Lambda$ systems.}}
Table \ref{SigmacSLambda} presents the bound state solutions for the $\Sigma_c^*\Lambda/\Sigma_c\Sigma/\Sigma_c^*\Sigma$ coupled systems with $1(0^-)$, $1(1^-)$, and $1(2^-)$. After solving the coupled channel Schr\"{o}dinger equations, we can see that  that the numerical results closely resemble those for the $\Sigma_c\Lambda$ systems as
\begin{itemize}
    \item In the single channel analysis, we cannot obtain the loosely bound state solutions for the single $\Sigma_c^*\Lambda$ systems with $1(0^-)$, $1(1^-)$, and $1(2^-)$.
    \item After considering the coupled channel effects, the OBE interactions become sufficiently strong to generate loosely bound states; however, no molecular candidates composed primarily of the $\Sigma_c^*\Lambda$ channels emerge..
    \item Due to the substantial probabilities, the $\Sigma_c^*\Lambda/\Sigma_c^*\Sigma$ and $\Sigma_c\Sigma$ channels play a significant role in binding the coupled states of $\Sigma_c^*\Lambda/\Sigma_c\Sigma/\Sigma_c^*\Sigma$ with $1(0^-, 2^-)$ and $1(1^-)$, respectively. This suggests that it may be possible to identify resonances arising from the $\Sigma_c^*\Lambda/\Sigma_c^*\Sigma$ coupled interactions with $1(0^-, 2^-)$ and the $\Sigma_c\Sigma$ interaction with $1(1^-)$. The corresponding results will be discussed in Section \ref{resonant}.
\end{itemize}

\renewcommand{\arraystretch}{2.0}
\begin{table}[!hbtp]
\renewcommand\tabcolsep{0.05cm}
\renewcommand{\arraystretch}{1.5}
\caption{The bound state solutions (the binding energy $E$, the root-mean-square radius $r_{RMS}$, and the probabilities $p_i$ for all the discussed channels) for the $\Sigma_c^*\Lambda$ systems with $I(J^P)=1(0^-)$, $1(1^-)$, and $1(2^-)$ after considered the coupled channel effects. Here, the unites for the cutoff $\Lambda$, the binding energy $E$, and the root-mean-square radius $r_{RMS}$ are GeV, MeV, and fm, respectively. }\label{SigmacSLambda}
\begin{tabular}{ccccccccccccc}
\toprule[1.5pt]
\toprule[1.5pt]
     $I(J^P)$                  &$\Lambda$    &$E$    &$r_{RMS}$    &$\Sigma_c^{*}\Lambda(^{3}P/^{5}P)$ &$\Sigma_c\Sigma(^{1}P/^{3}P)$        &$\Sigma_c^{*}\Sigma(^{3}P/^{5}P)$  \\  \hline
    $1(0^-)$        &1.287     &$-$1.94        &0.96           &{\bf{46.03}}/--   &--/3.41    &{\bf{50.55}}/--    \\
                    &1.290     &$-$4.48        &0.81           &{\bf{44.56}}/--   &--/3.58      &{\bf{51.86}}/--     \\
                    &1.293     &$-$7.15        &0.74           &{\bf{43.58}}/--     &--/3.72     &{\bf{52.70}}/--       \\ \hline
     $1(1^-)$     &1.560     &$-$2.61        &0.82        &1.59/22.80   &{\bf{64.41}}/5.76    &0.27/5.14    \\
         & 1.565     &$-$6.26        &0.75       &1.53/23.21   &{\bf{63.68}}/5.75      &0.29/5.55  \\
                &1.570     &$-$10.14        &0.70  &1.51/23.72     &{\bf{62.83}}/5.70     &0.30/5.94     \\ \hline
     $1(2^-)$  &1.287     &$-$4.37       &0.83    &{\bf{38.23}}/2.03   &--/16.13      &{\bf{39.05}}/4.54     \\
      &1.290     &$-$7.14        &0.75    &{\bf{37.77}}/1.65   &--/16.17    &{\bf{39.97}}/4.42    \\

             & 1.293     &$-$10.04        &0.71       &{\bf{37.47}}/1.38     &--/16.16     &{\bf{40.70}}/4.30  \\
    \bottomrule[1.5pt]
\bottomrule[1.5pt]
\end{tabular}
\end{table}

\paragraph{\bf{The $\Sigma_c\Sigma$ systems.}}
In Table \ref{SigmacSigma}, we collect the bound state solutions for the $\Sigma_c\Sigma$ systems with $I(J^P)=0(0^-)$,$0(1^-)$,$0(2^-)$,$1(0^-)$, $1(1^-)$, and $1(2^-)$. In the single channel case, the $\Sigma_c\Sigma$ state with $0(1^-)$ can be prime molecular candidate, as we can obtain the loosely bound state solutions with $\Lambda$ around 1.00 GeV and the RMS radius is around 1 fm or larger. For the remaining $\Sigma_c\Sigma$ systems, the OBE interactions are insufficiently strong to generate loosely bound states.

In the coupled channel case, we agian cannot obtain the loosely bound state solutions for the $\Sigma_c\Sigma/\Sigma_c^*\Sigma_c$ coupled systems with $1(0^-)$ and $1(2^-)$. However, the $\Sigma_c\Sigma/\Sigma_c^*\Sigma$ coupled molecular state with quantum number $1(1^-)$ binds slightly deeper than the corresponding single channel state. The dominant channel in this case is $\Sigma_c\Sigma|{}^5P_1\rangle$. Compared to the $\Sigma_c\Sigma/\Sigma_c^*\Sigma$ coupled molecular state with $0(1^-)$, the OBE interactions are somewhat weaker due to the larger cutoff value. Additionally, the coupled channel effects play a crucial role in the formation of this molecular candidate, as evidenced by the shift in the binding energy and structure compared to the single channel case.

For the $\Sigma_c\Sigma/\Sigma_c^*\Sigma_c$ coupled systems with $0(0^-)$ and $0(2^-)$, although loosely bound state solutions appear at a cutoff value around 1.00 GeV, these systems do not qualify as good molecular candidates. The primary reason is that the dominant channels are not the $\Sigma_c\Sigma$ states, but rather the $\Sigma_c^*\Sigma$ channels, which are associated with higher masses. This results in much smaller RMS radii, around 0.5 fm, which is inconsistent with the typical size of hadronic molecules.

\renewcommand{\arraystretch}{2.0}
\begin{table*}[!hbtp]
\renewcommand\tabcolsep{0.6cm}
\renewcommand{\arraystretch}{1.5}
\caption{The bound state solutions (the binding energy $E$, the root-mean-square radius $r_{RMS}$, and the probabilities $p_i$ for all the discussed channels) for the $\Sigma_c\Sigma$ systems with $I(J^P)=0(0^-)$, $0(1^-)$, $0(2^-)$, $1(0^-)$, $1(1^-)$, and $1(2^-)$. Here, the unites for the cutoff $\Lambda$, the binding energy $E$, and the root-mean-square radius $r_{RMS}$ are GeV, MeV, and fm, respectively. }\label{SigmacSigma}
\begin{tabular}{ccccccccccccc}
\toprule[1.5pt]
\toprule[1.5pt]
Single channel &$I(J^P)$  &$\Lambda$    &$E$    &$r_{RMS}$   &$\Sigma_c\Sigma(^{3}P/^{5}P)$
       \\ \cline{2-6}
&$0(1^-)$   &1.100    &$-$0.10        &2.86          &100/0.00              \\
                      && 1.110    &$-$4.57        &1.19         &100/0.00
   \\
                       &&1.120    &$-$9.84        &1.00         &100/0.00
  \\ \midrule[1.5pt]
Couple channel &$I(J^P)$                  &$\Lambda$    &$E$    &$r_{RMS}$    &$\Sigma_c\Sigma(^{3}P/^{5}P)$        &$\Sigma_c^{*}\Sigma(^{3}P/^{5}P)$  \\ \cline{2-7}
    &$0(0^-)$        & 1.150     &$-$2.90        &0.68                     &--/1.63   &{\bf{98.36}}/--         \\
    &&1.155     &$-$7.47        &0.67                    &--/1.70   &{\bf{98.30}}/--        \\
                       & &1.160     &$-$12.18        &0.65                     &--/1.76   &{\bf{98.23}}/--         \\
                         \cline{2-7}
     &$0(1^-)$     &0.930     &$-$0.04        &5.15                    &{\bf{29.25/68.48}}   &0.37/1.89         \\

                       && 0.950     &$-$4.90        &1.65                    &{\bf{38.55/57.89}}   &0.72/2.82       \\
                        & &0.970     &$-$11.83        &1.30                     &{\bf{43.64/52.22}}   &1.05/3.07         \\\cline{2-7}
     &$0(2^-)$    & 1.165     &$-$2.19        &0.77                     &--/13.70   &{\bf{82.27}}/4.03         \\
     && 1.170     &$-$6.89        &0.69                   &--/12.71   &{\bf{83.44}}/3.83         \\

                        &&1.175     &$-$11.79        &0.67                    &--/12.21   &{\bf{84.18}}/3.62  \\\cline{2-7}
    &$1(0^-)$        &--/--      &--/--         &--/--                      &--/--   &--/--         \\
                               \cline{2-7}
     &$1(1^-)$     &1.850     &$-$0.37        &2.80                    &5.41/{\bf{91.23}}      &0.22/3.12         \\
                        &&1.900     &$-$3.68        &1.57                     &6.25/{\bf{89.31}}       &0.24/4.18        \\
                      & &1.950     &$-$7.86        &1.28                    &6.81/{\bf{88.00}}       &0.23/4.94       \\ \cline{2-7}
    & $1(2^-)$    &--/--      &--/--         &--/--                      &--/--   &--/--   \\
    \bottomrule[1.5pt]
\bottomrule[1.5pt]
\end{tabular}
\end{table*}

\paragraph{\bf{The $\Sigma_c^*\Sigma$ systems.}}
In the cutoff range of $0.80\leq\Lambda\leq2.00$ GeV, we can obtain the loosely bound state solutions for the $\Sigma_c^*\Sigma$ systems with $0(0^-, 1^-, 2^-)$ and $1(2^-)$ as shown in Table \ref{SigmacSSigma}. Compared to the isoscalar bound states, the isovector bound state exhibits slightly shallower binding, primarily due to a slightly larger cutoff value. The low-spin components of the $\Sigma_c^*\Sigma$ channels play a significant role in the formation of low-spin bound states, such as the $\Sigma_c^*\Sigma$ states with quantum numbers $0(0^-)$ and $0(1^-)$. In contrast, the $\Sigma_c^*\Sigma|{}^5P_2\rangle$ channel contributes significantly to the formation of bound states with quantum numbers $0(2^-)$ and $1(2^-)$.

For the remaining $\Sigma_c^*\Sigma$ systems with quantum numbers $1(0^-)$ and $1(1^-)$, no loosely bound state solutions are found for cutoffs $\Lambda < 2.00$ GeV. This indicates that the OBE effective potentials are not strong enough to generate sufficient attractive interactions in these channels.

Based on the reasonable cutoff values and the corresponding loosely bound state solutions, we conclude that four potential molecular candidates exist: the $\Sigma_c^*\Sigma$ molecules with quantum numbers $0(0^-)$, $0(1^-)$, $0(2^-)$, and $1(2^-)$.

\renewcommand{\arraystretch}{2.0}
\begin{table}[!hbtp]
\renewcommand\tabcolsep{0.4cm}
\renewcommand{\arraystretch}{1.5}
\caption{The bound state solutions (the binding energy $E$, the root-mean-square radius $r_{RMS}$, and the probabilities $p_i$ for all the discussed channels) for the $\Sigma_c^*\Sigma$ systems with $I(J^P)=0(0^-)$, $0(1^-)$, $0(2^-)$, $1(0^-)$, $1(1^-)$, and $1(2^-)$. Here, the unites for the cutoff $\Lambda$, the binding energy $E$, and the root-mean-square radius $r_{RMS}$ are GeV, MeV, and fm, respectively. }\label{SigmacSSigma}
\begin{tabular}{ccccc}
\toprule[1.5pt]
\toprule[1.5pt]
$I(J^P)$  &$\Lambda$    &$E$    &$r_{RMS}$   &
$\Sigma_c^{*}\Sigma(^{3}P/^{5}P)$\\ \hline
 $0(0^-)$  &1.050    &$-$1.15        &1.98   &\multicolumn{1}{c}{100/--} \\
                            &1.060    &$-$3.95        &1.46   &\multicolumn{1}{c}{100/--} \\
                        & 1.070    &$-$7.21        &1.26   &\multicolumn{1}{c}{100/--}   \\ \hline
$0(1^-)$   &1.130    &$-$0.77        &2.16   &\multicolumn{1}{c}{80.22/19.77}  \\
                         &1.140    &$-$3.67        &1.46  &\multicolumn{1}{c}{82.44/17.55}  \\
                       & 1.150    &$-$7.13        &1.23   &\multicolumn{1}{c}{83.93/16.06}  \\\hline
$0(2^-)$ &1.060    &$-$0.42        &3.29   &34.01/65.98         \\
         &1.090    &$-$4.68        &1.72   &47.41/52.58        \\
         & 1.120    &$-$11.74        &1.28   &60.30/39.69               \\ \hline
$1(2^-)$  &1.800  &$-$0.80  &2.41  &24.59/75.40     \\
           &1.850  &$-$3.76  &1.62  &25.73/74.26          \\
           &1.900  &$-$7.42  &1.35  &26.35/73.64           \\
\bottomrule[1.5pt]
\bottomrule[1.5pt]
\end{tabular}
\end{table}

All in all, the current results suggest the existence of several potential molecular candidates, which include the $\Xi_c^{\prime}N$ molecule with $I(J^P)=0(1^-)$, the $\Xi_c^{*}N$ molecules with $0(0^-)$, $0(1^-)$, $0(2^-)$, the $\Sigma_c\Sigma$ molecules with $0(1^-)$ and $1(1^-)$, the $\Sigma_c^*\Sigma$ molecules with $0(0^-)$, $0(1^-)$, $0(2^-)$, and $1(2^-)$. Here, we also find the coupled channel effects play an important role  in the formation of the $\Sigma_c\Sigma$ molecule with $1(1^-)$.

\subsection{Resonances}\label{resonant}

In the previous subsection, we not only identified the existence of several potential molecular candidates, but also observed that the OBE interactions from other coupled channel systems can be sufficiently strong, thereby motivating the search for possible resonances. The corresponding coupled channel systems are summarized as follows,
\begin{eqnarray*}
0(0^-):    &&\Xi_cN/\Xi_c^{*}N,\quad  \Xi_c^*N/\Sigma_c^*\Sigma,\\
0(1^-):    &&\Xi_cN/\Xi_c^{\prime}N/\Xi_c^*N,    \\
0(2^-):    &&\Xi_cN/\Xi_c^*N,\quad    \Xi_c^{*}N/\Sigma_c^*\Sigma,\\
1(0^-):    &&\Lambda_c\Sigma/\Sigma_c^*\Lambda/\Sigma_c^*\Sigma,
              \Sigma_c^*\Lambda/\Sigma_c^*\Sigma,\\
1(1^-):    &&\Lambda_c\Sigma/\Sigma_c\Lambda/\Sigma_c\Sigma,
              \Sigma_c\Lambda/\Sigma_c\Sigma,\\
1(2^-):    &&\Lambda_c\Sigma/\Sigma_c^*\Lambda/\Sigma_c^*\Sigma,\quad
              \Sigma_c^*\Lambda/\Sigma_c^*\Sigma.
\end{eqnarray*}

As is well known, the resonances appear as the phase shifts equal $\delta(E_r)=(n+1/2)\pi$ with $n=0,1,2,\cdots$ This leads to the maximum of the scattering cross section, $\sigma_t=\frac{4\pi}{2\mu E}\sum_{l=0}^{\infty}(2l+1)\text{sin}^2\delta_l(E)$. The resonant width is defined as $\Gamma_r=2/\left(\frac{d\delta}{dE}\right)_{E_r}$.

\begin{figure}[!htbp]
\centering
$\left.\begin{array}{ll}
\includegraphics[width=1.6in]{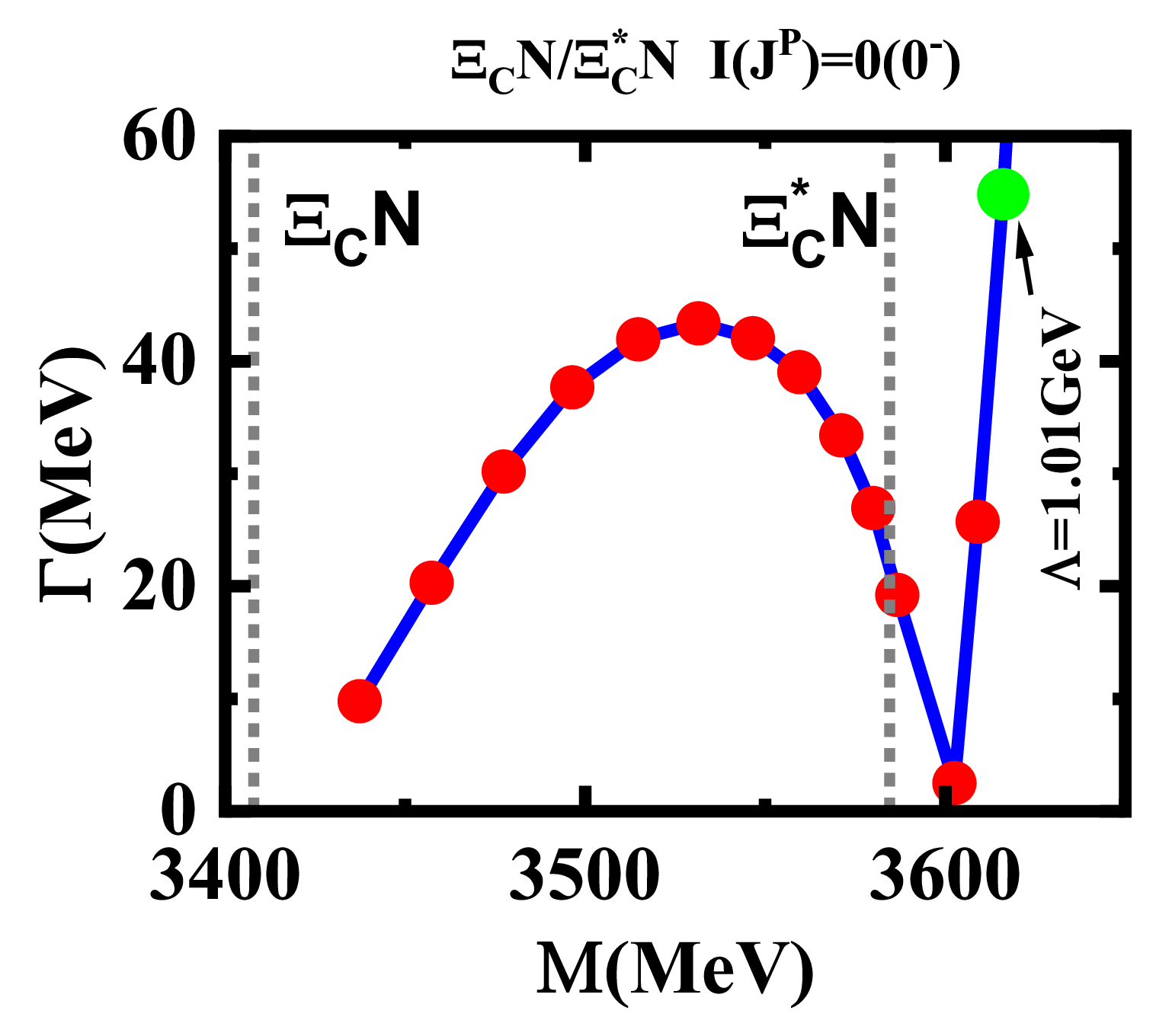}
&\includegraphics[width=1.6in]{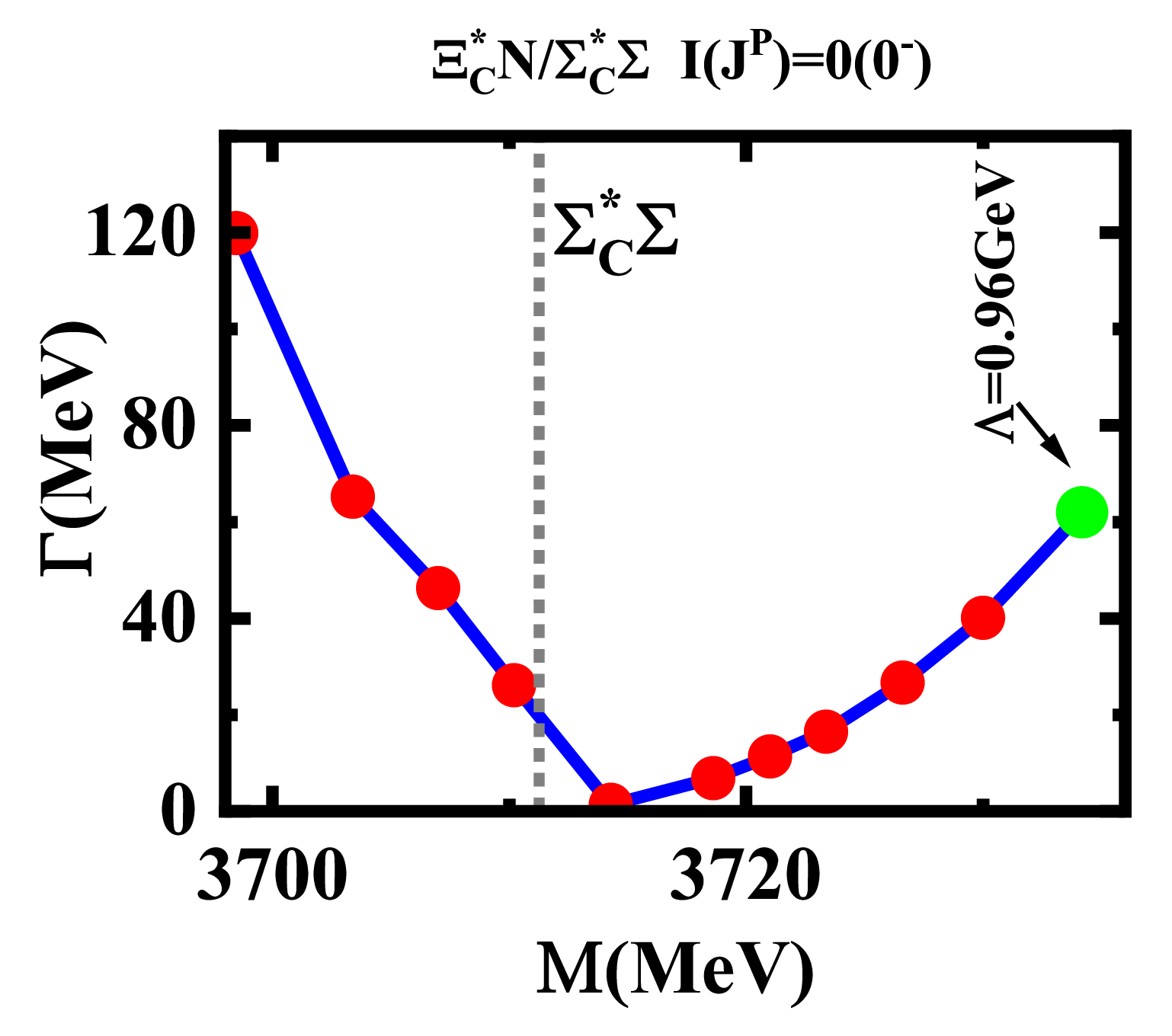}\\
\includegraphics[width=1.6in]{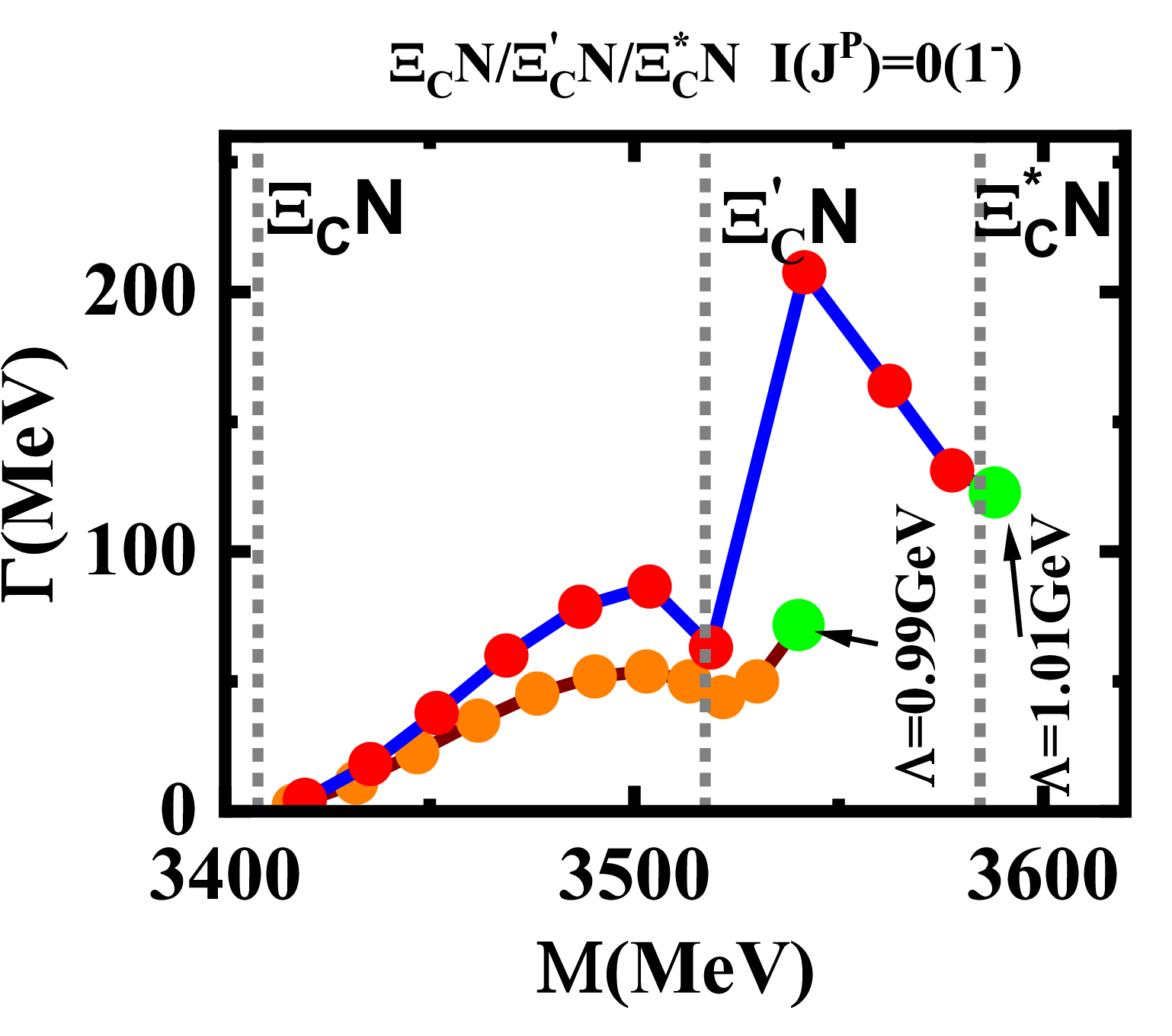}\\
\includegraphics[width=1.6in]{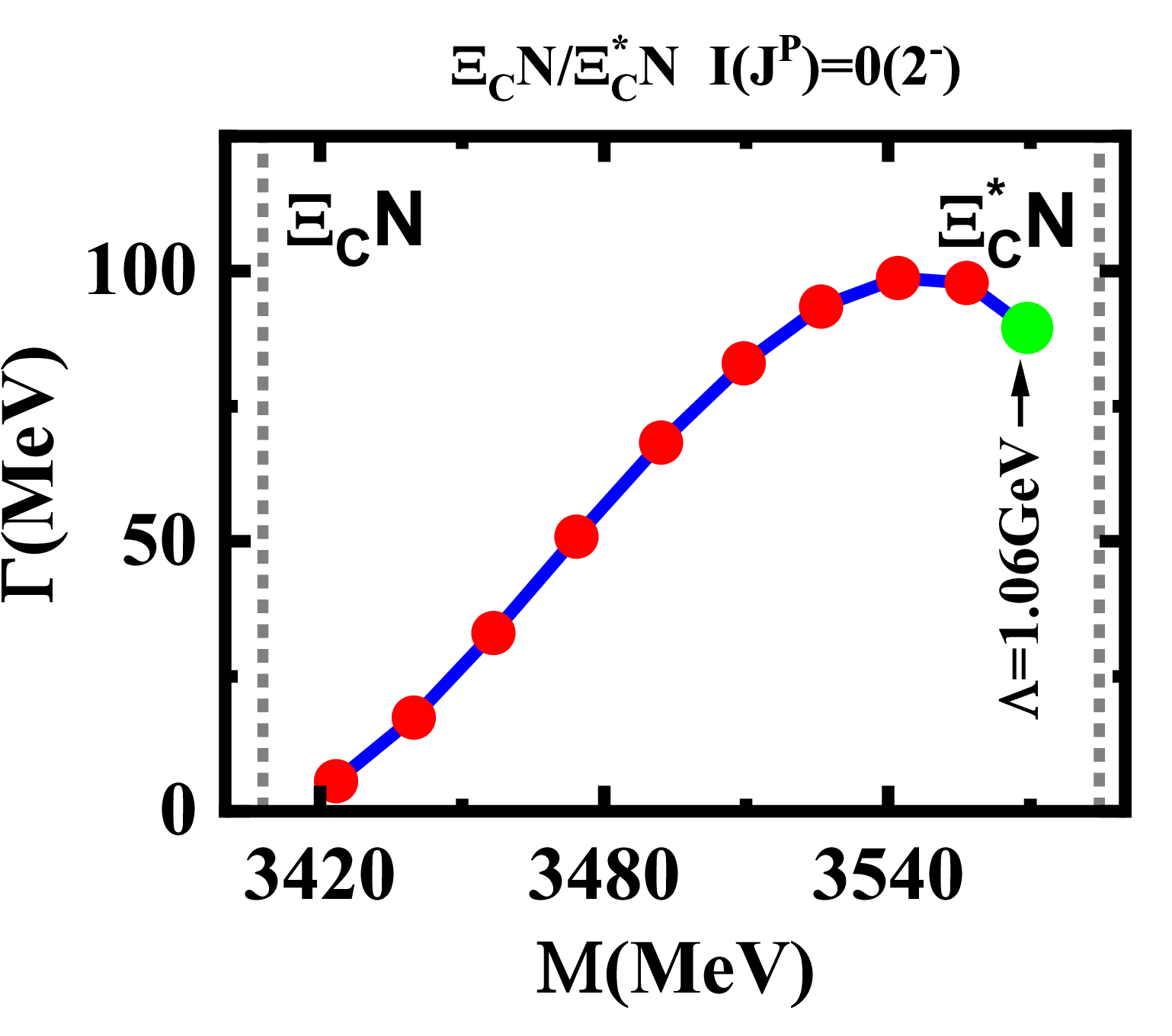}
&\includegraphics[width=1.6in]{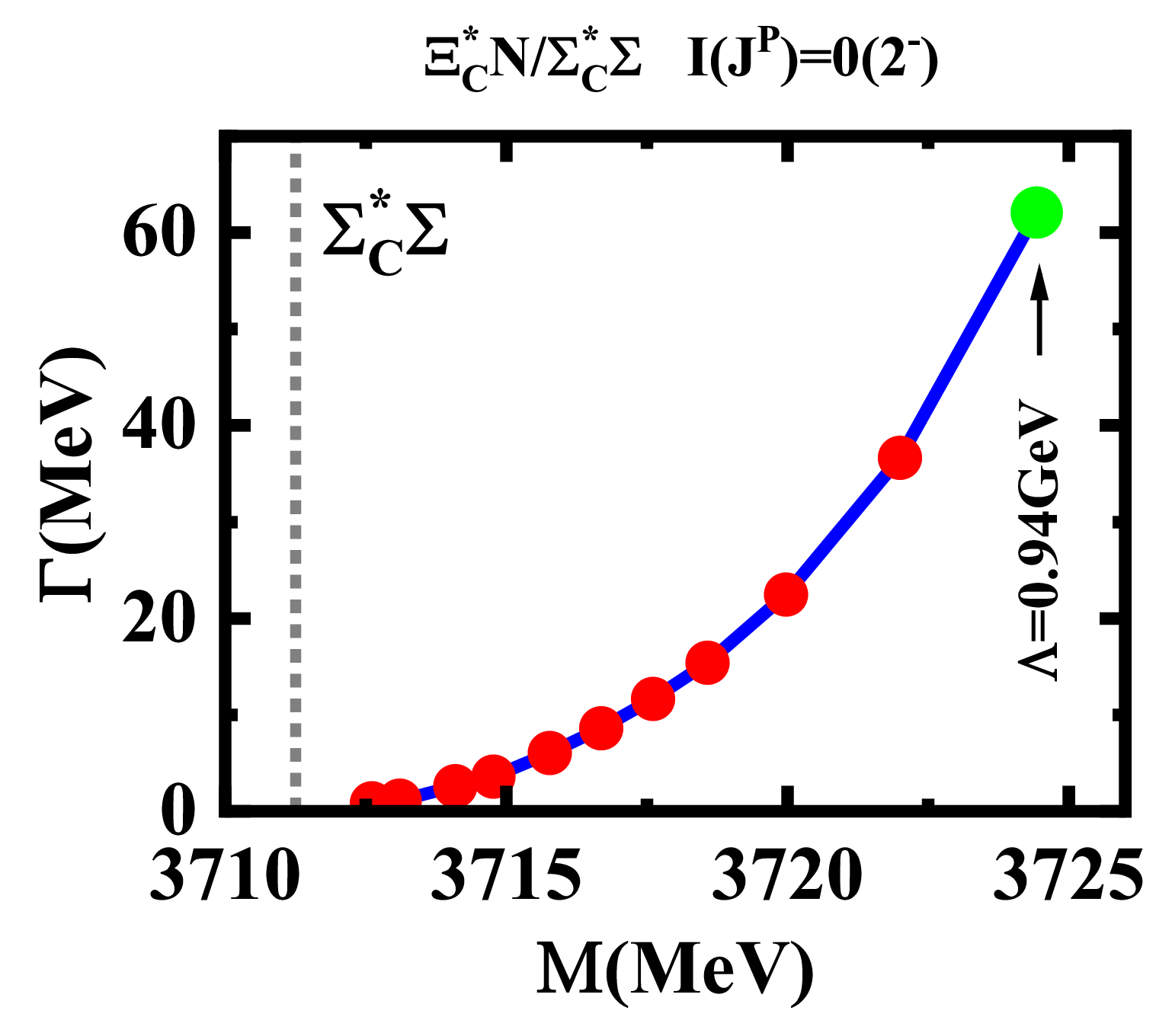}\\
\includegraphics[width=1.6in]{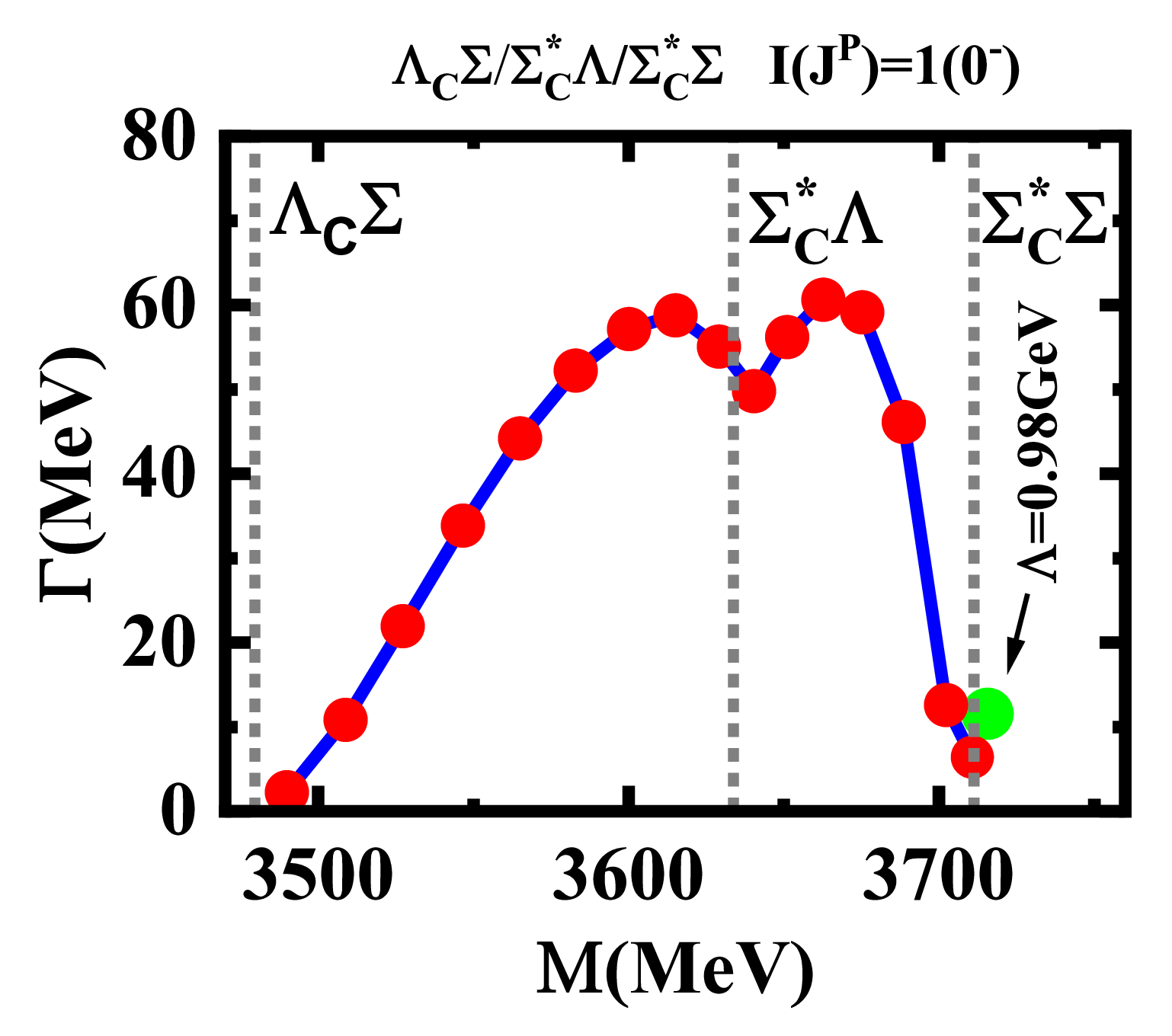}
&\includegraphics[width=1.6in]{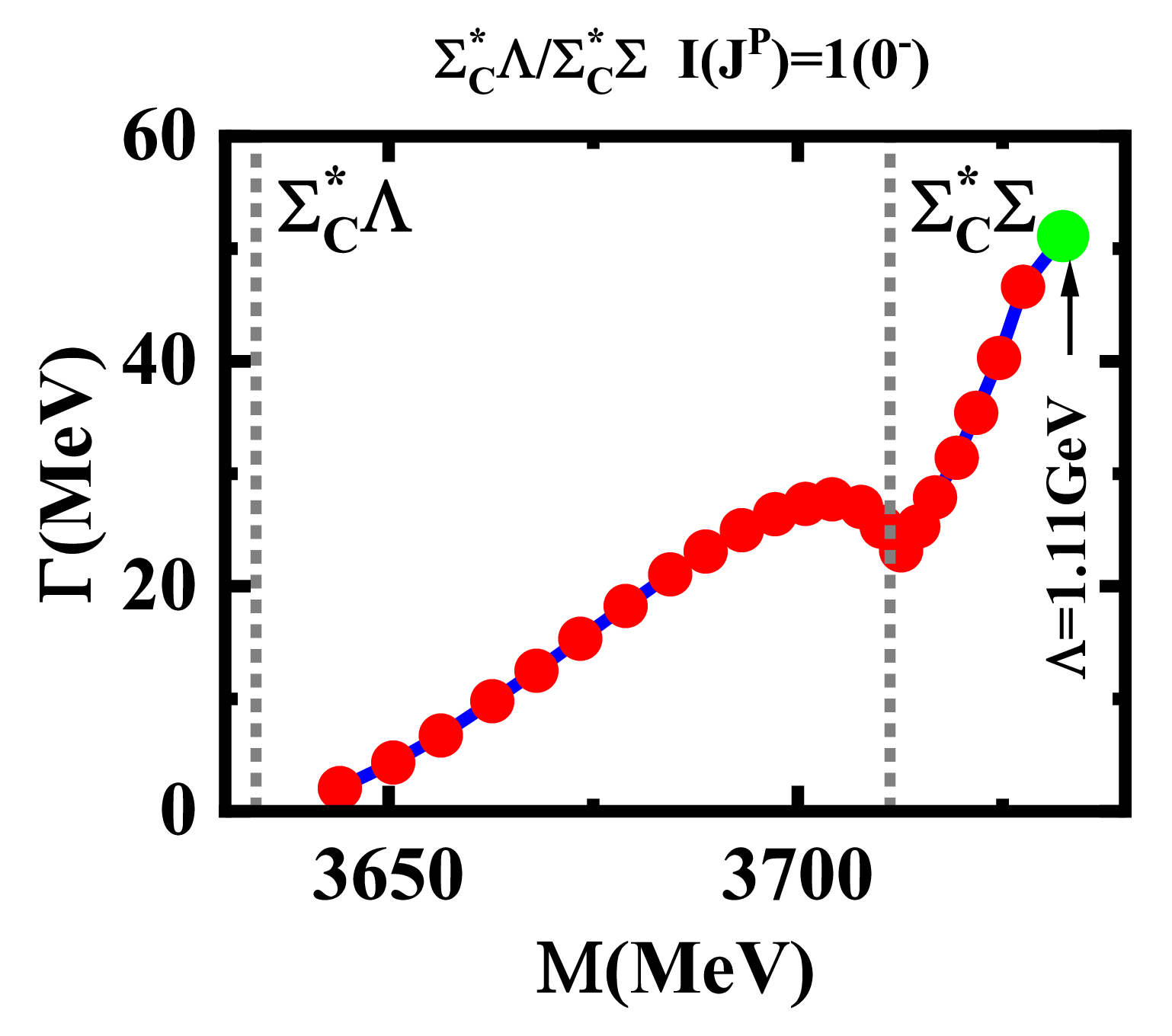}\\
\includegraphics[width=1.6in]{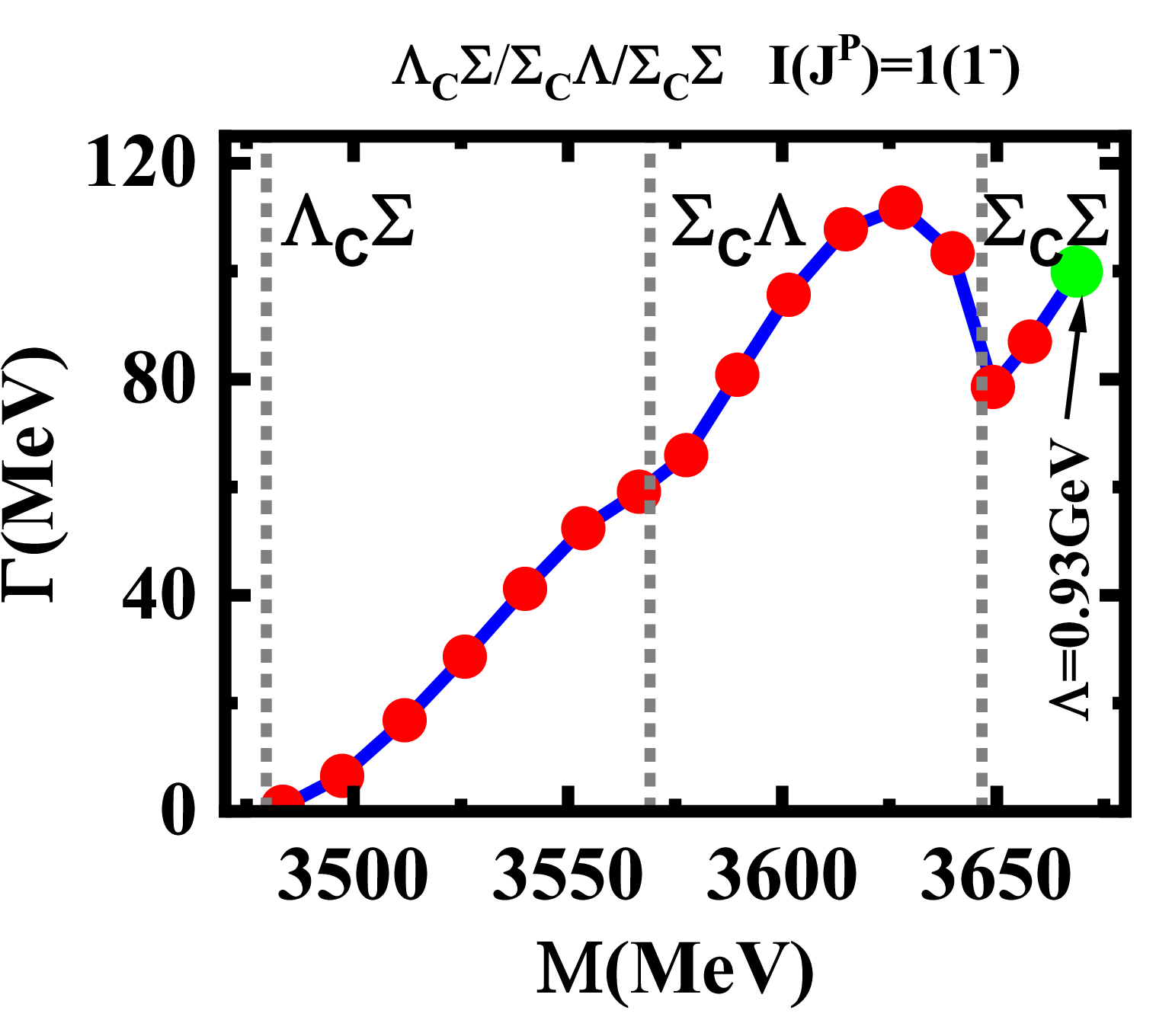}
&\includegraphics[width=1.6in]{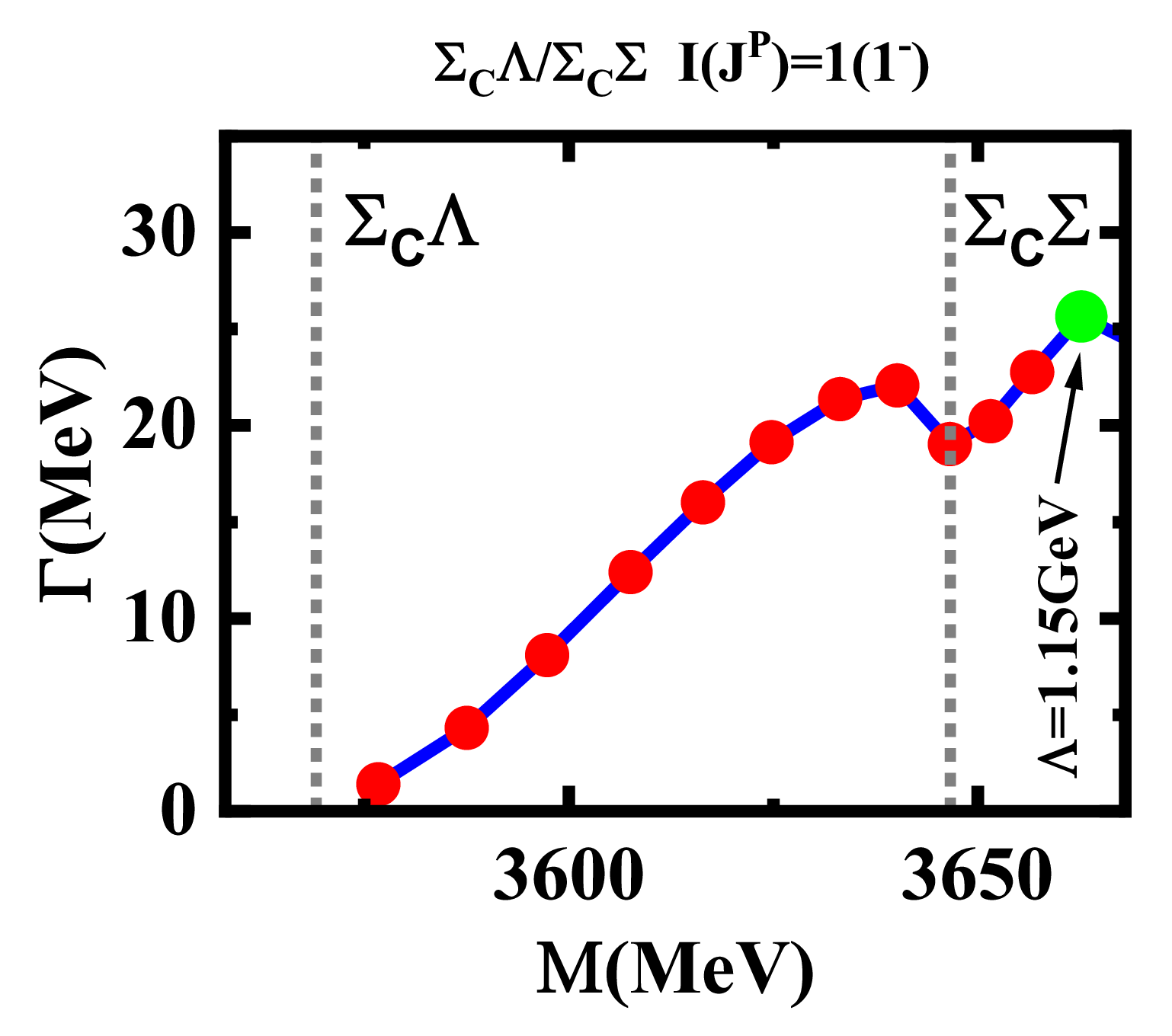}\\
\includegraphics[width=1.6in]{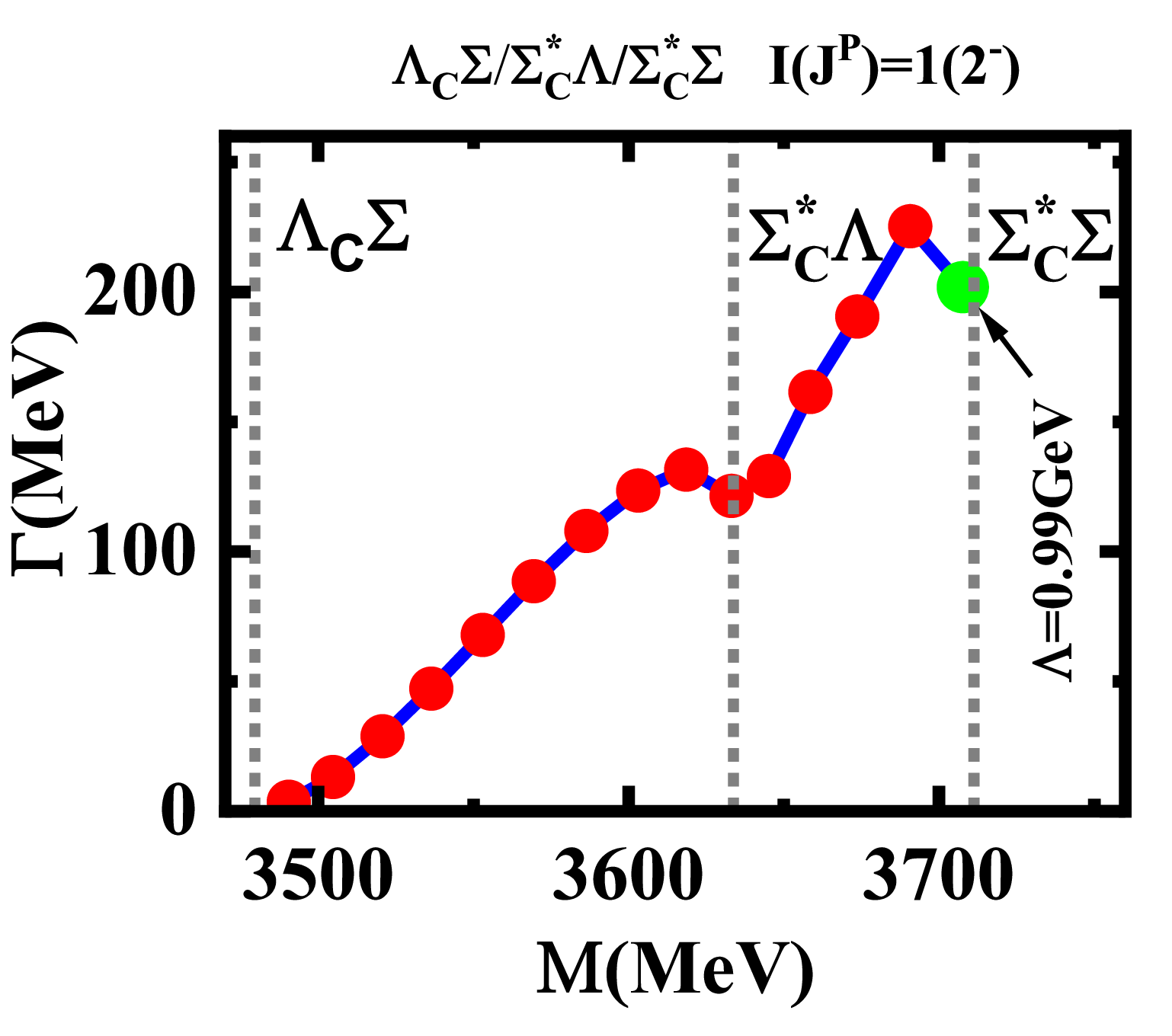}
&\includegraphics[width=1.6in]{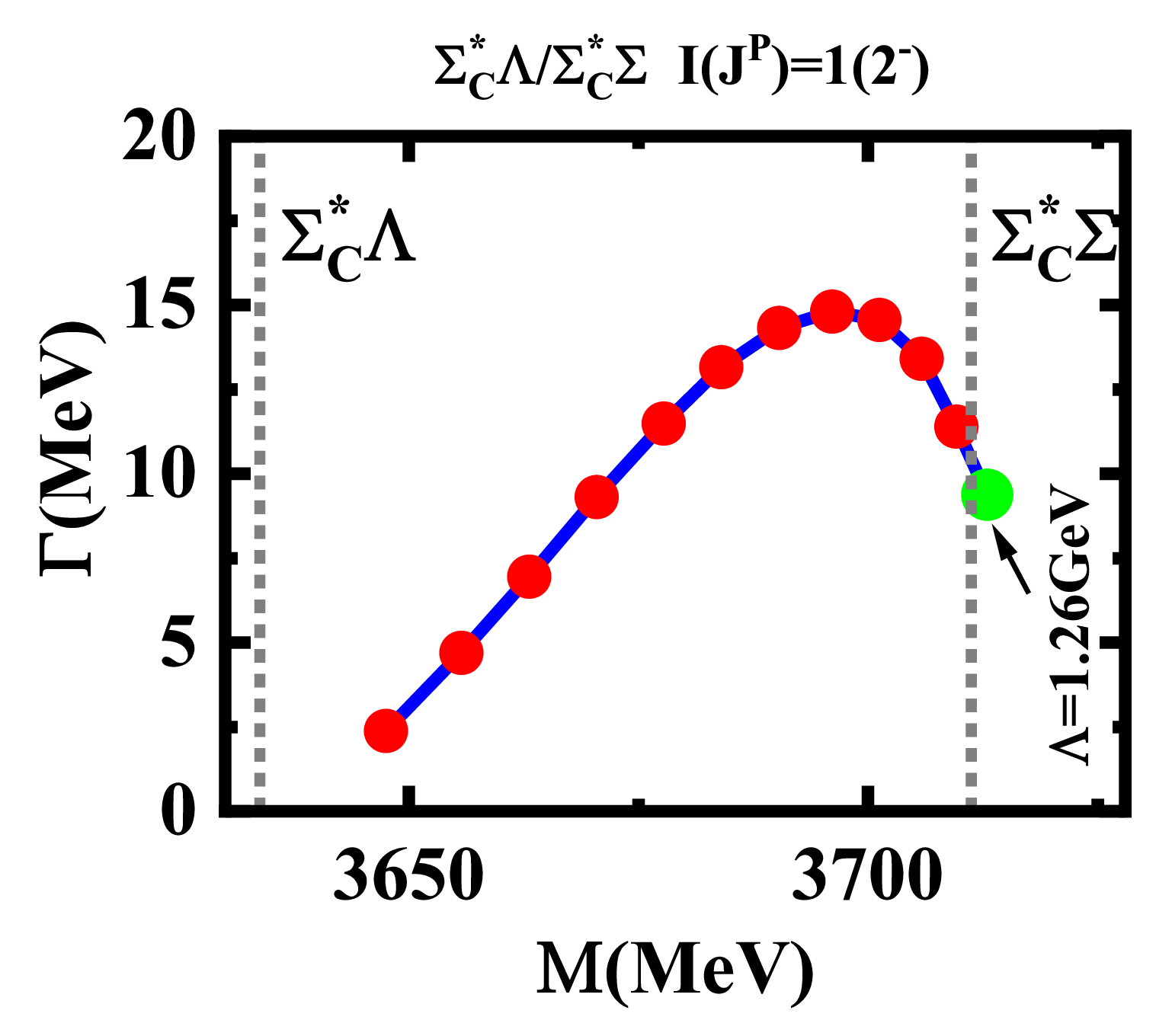}\end{array}\right.$
\caption{The cutoff dependence of the obtained resonant masses and decay widths from the discussed coupled channels systems. Here, the green dots represent the cutoff values when the resonances firstly appear. And the cutoff gap between two adjacent points is set as $\Delta \Lambda=0.01$ GeV.}
\label{r1}
\end{figure}

In Figure \ref{r1}, we present the relevant resonant parameters. For all the discussed channels, resonances can be obtained with a cutoff around 1.00 GeV, which further confirms that the corresponding OBE effective potentials are strong enough, as discussed in the previous subsection \ref{bound}. Specifically, we observe the following situations,
\begin{enumerate}
    \item \textbf{$\Xi_cN/\Xi_c^*N$ interactions with $0(0^-)$:} A resonance appears with a mass above the $\Xi_c^*N$ threshold, and the decay width is on the order of several tens of MeV. As the cutoff value increases, the OBE interactions become stronger, which leads to a deeper binding and reduces the mass of the resonance below the $\Xi_c^*N$ threshold. Compared to the bound state solution for the $\Xi_c^*N$ molecule with $0(0^-)$, the slightly smaller cutoff values here result in weaker OBE effective potentials, causing the bound state evolves into a resonance. Additionally, the evolution of the decay width with the cutoff shows an inflection point at the $\Xi_c^*N$ threshold, further indicating the close relationship between this resonance and the $\Xi_c^*N$ channel. In conclusion, this resonance is not an independent state but corresponds to the $\Xi_c^*N$ molecule with $0(0^-)$. Although no new state appears in the coupled channel phase shift analysis, these results provide important information on the total decay width of the $\Xi_c^*N$ bound state.

    \item \textbf{$\Xi_c^*N/\Sigma_c^*\Sigma$ coupled interactions with $0(0^-)$:} A resonance is obtained with a mass near the $\Sigma_c^*\Sigma$ threshold, where the cutoff is approximately 1.00 GeV. The cutoff dependence of the resonant parameters mirrors that of the $\Xi_cN/\Xi_c^*N$ interactions with $0(0^-)$, where the mass decreases with increasing cutoff, and the decay width exhibits a minimum near the $\Sigma_c^*\Sigma$ threshold. This behavior suggests a close relationship between this resonance and the predicted $\Sigma_c^*\Sigma$ molecule with $0(0^-)$, as shown in Table \ref{SigmacSSigma}.

    \item \textbf{$\Xi_cN/\Xi_c^{\prime}N/\Xi_c^*N$ coupled interactions with $0(1^-)$:} Two resonances can be observed at a cutoff of 1.01 GeV. The broad resonance, with a width exceeding one hundred MeV, is located near the $\Xi_c^*N$ threshold, while the narrow resonance, with a width around fifty MeV, is near the $\Xi_c^{\prime}N$ threshold. The cutoff dependence of the decay widths indicates that the broad and narrow resonances are associated with the $\Xi_c^*N$ and $\Xi_c^{\prime}N$ channels, respectively. After comparing the bound state solutions for the $\Xi_c^{\prime}N/\Sigma_c\Lambda/\Xi_c^*N/\Sigma_c^*\Lambda/\Sigma_c\Sigma/\Sigma_c^*\Sigma$ coupled systems with $0(1^-)$ in Table \ref{XicpN} and the $\Xi_c^*N/\Sigma_c^*\Lambda/\Sigma_c\Sigma/\Sigma_c^*\Sigma$ coupled systems with $0(1^-)$ in Table \ref{XicSN}, it becomes clear that these two resonances are not new structures, but correspond to the $\Xi_c^{\prime}N$ and $\Xi_c^*N$ molecular states predicted earlier.

    \item \textbf{$\Xi_cN/\Xi_c^*N$ coupled interactions with $0(2^-)$:} A resonance appears in the reasonable cutoff region, and this state is closely related to the $\Xi_c^*N$ molecular state with $0(2^-)$ predicted in Table \ref{XicSN}. The decay width is on the order of one hundred MeV.

    \item \textbf{$\Xi_c^*N/\Sigma_c^*\Sigma$ coupled interactions with $0(2^-)$:} A resonance emerges with a mass near the $\Sigma_c^*\Sigma$ threshold when the cutoff is around 1.00 GeV. This state is closely related to the $\Sigma_c^*\Sigma$ molecular state with $0(2^-)$, as predicted in Table \ref{SigmacSSigma}.

    \item \textbf{$\Lambda_c\Sigma/\Sigma_c^*\Lambda/\Sigma_c^*\Sigma$ coupled interactions with $1(0^-)$:} A resonance can be obtained with a mass around the $\Sigma_c^*\Sigma$ threshold, and the decay width is on the order of several tens of MeV. As the cutoff increases, two inflection points of the decay width appear at the $\Sigma_c^*\Sigma$ and $\Sigma_c^*\Lambda$ thresholds, implying a close relation between the channels. Further analysis on the single $\Sigma_c^*\Sigma$ channel also yields a resonance with similar parameters. Therefore, this state is a shape-type resonance dominated by the $\Sigma_c^*\Sigma$ channel. Given that no loosely bound state solutions for the $\Sigma_c^*\Sigma$ molecule with $1(0^-)$ are found in the reasonable cutoff range (as shown in Table \ref{SigmacSSigma}), we conclude that the coupled-channel effects play a key role in the formation of this shape-type resonance.

    \item \textbf{$\Sigma_c^*\Lambda/\Sigma_c^*\Sigma$ coupled interactions with $1(0^-)$:} The results are similar to those in the $\Lambda_c\Sigma/\Sigma_c^*\Lambda/\Sigma_c^*\Sigma$ coupled interactions with $1(0^-)$, and we predict the existence of a shape-type $\Sigma_c^*\Sigma$ resonance with $1(0^-)$. The slightly larger cutoff value compared to that in the $\Lambda_c\Sigma/\Sigma_c^*\Lambda/\Sigma_c^*\Sigma$ coupled interactions suggests that the $\Lambda_c\Sigma$ channel plays a positive role in generating this shape-type resonance.

    \item \textbf{$\Lambda_c\Sigma/\Sigma_c\Lambda/\Sigma_c\Sigma$ coupled interactions with $1(1^-)$:} A resonance appears in a reasonable cutoff region, with the $\Sigma_c\Sigma$ channel playing an important role. The inflection point of the decay width occurs near the $\Sigma_c\Sigma$ threshold. A further check of the phase shifts of the $\Lambda_c\Sigma/\Sigma_c\Sigma$ coupled interactions, the $\Sigma_c\Lambda/\Sigma_c\Sigma$ coupled interactions, and the single $\Sigma_c\Sigma$ interactions with $1(1^-)$ shows that the cutoff dependence of the resonant parameters for the $\Lambda_c\Sigma/\Sigma_c\Sigma$ coupled interactions and the $\Lambda_c\Sigma/\Sigma_c\Lambda/\Sigma_c\Sigma$ coupled interactions are very similar. This indicates that the resonance is a Feshbach-type resonance, where both the $\Lambda_c\Sigma$ and $\Sigma_c\Sigma$ channels play important roles.

    \item \textbf{$\Sigma_c\Lambda/\Sigma_c\Sigma$ coupled interactions with $1(1^-)$:} A resonance appears in a reasonable cutoff region, with the $\Sigma_c\Sigma$ channel playing a crucial role. The phase shift analysis for the single $\Sigma_c\Sigma$ channel with $1(1^-)$ shows a resonance near the $\Sigma_c\Sigma$ threshold, and the obtained resonant parameters are very similar to those for the $\Sigma_c\Lambda/\Sigma_c\Sigma$ coupled interactions with $1(1^-)$. Thus, this resonance is a shape-type resonance.

    \item \textbf{$\Lambda_c\Sigma/\Sigma_c^*\Lambda/\Sigma_c^*\Sigma$ coupled interactions with $1(2^-)$:} A resonance appears in the reasonable cutoff region. After comparing the phase shifts under the $\Lambda_c\Sigma/\Sigma_c^*\Sigma$ coupled interactions, the single $\Sigma_c\Sigma$ channel interactions, and the $\Lambda_c\Sigma/\Sigma_c^*\Lambda/\Sigma_c^*\Sigma$ coupled interactions with $1(2^-)$, we find that the resonant parameters obtained in these two coupled channel interactions are very close. This suggests that the resonance is a Feshbach-type resonance, with the $\Lambda_c\Sigma$ and $\Sigma_c^*\Sigma$ channels playing key roles.

    \item \textbf{$\Sigma_c^*\Lambda/\Sigma_c^*\Sigma$ coupled interactions with $1(2^-)$:} A resonance is obtained with a cutoff around 1.00 GeV. This state is closely related to the $\Sigma_c^*\Sigma$ molecule with $1(2^-)$. Since the cutoff here is slightly smaller than that for the molecular state predicted in Table \ref{SigmacSSigma}, the coupled channel effects play a positive role in the formation of this resonance.
\end{enumerate}

In conclusion, after analyzing the phase shifts of the $P$-wave interactions between a charmed baryon and a light baryon, we have not only confirmed the predictions of potential molecular candidates discussed in Section \ref{bound}, but also proposed the existence of several resonances. These resonances include the $\Sigma_c\Sigma$ shape-type resonance with $1(1^-)$, the $\Sigma_c^*\Sigma$ shape-type resonance with $1(0^-)$, the $\Lambda_c\Sigma/\Sigma_c\Sigma$ coupled Feshbach-type resonance with $1(1^-)$, and the $\Lambda_c\Sigma/\Sigma_c^*\Sigma$ coupled Feshbach-type resonance with $1(0^-)$ and $1(2^-)$. These findings offer valuable insights into the interactions between charmed and light baryons, and will further our understanding of their possible molecular and resonant structures.

\section{Summary}\label{sec4}

With the advancement of experimental precision, studies of hadron spectroscopy has entered into a new era. Theoretical investigations into hadron interactions serve as a powerful tool for deepening our understanding of hadron spectroscopy, while also offering valuable insights into the exploration of the complex, non-perturbative dynamics in QCD. In this work, we systematically examine the $P$-wave interactions between charmed and light baryons using the OBE model, considering the effects of coupled channels. After solving the coupled channel Schr\"{o}dinger equations and analyzing the resulting phase shifts, we predict the spectroscopic properties of charm-strange di-baryon structures with negative parity.

\begin{figure}[!htbp]
\centering
\includegraphics[width=3.4in]{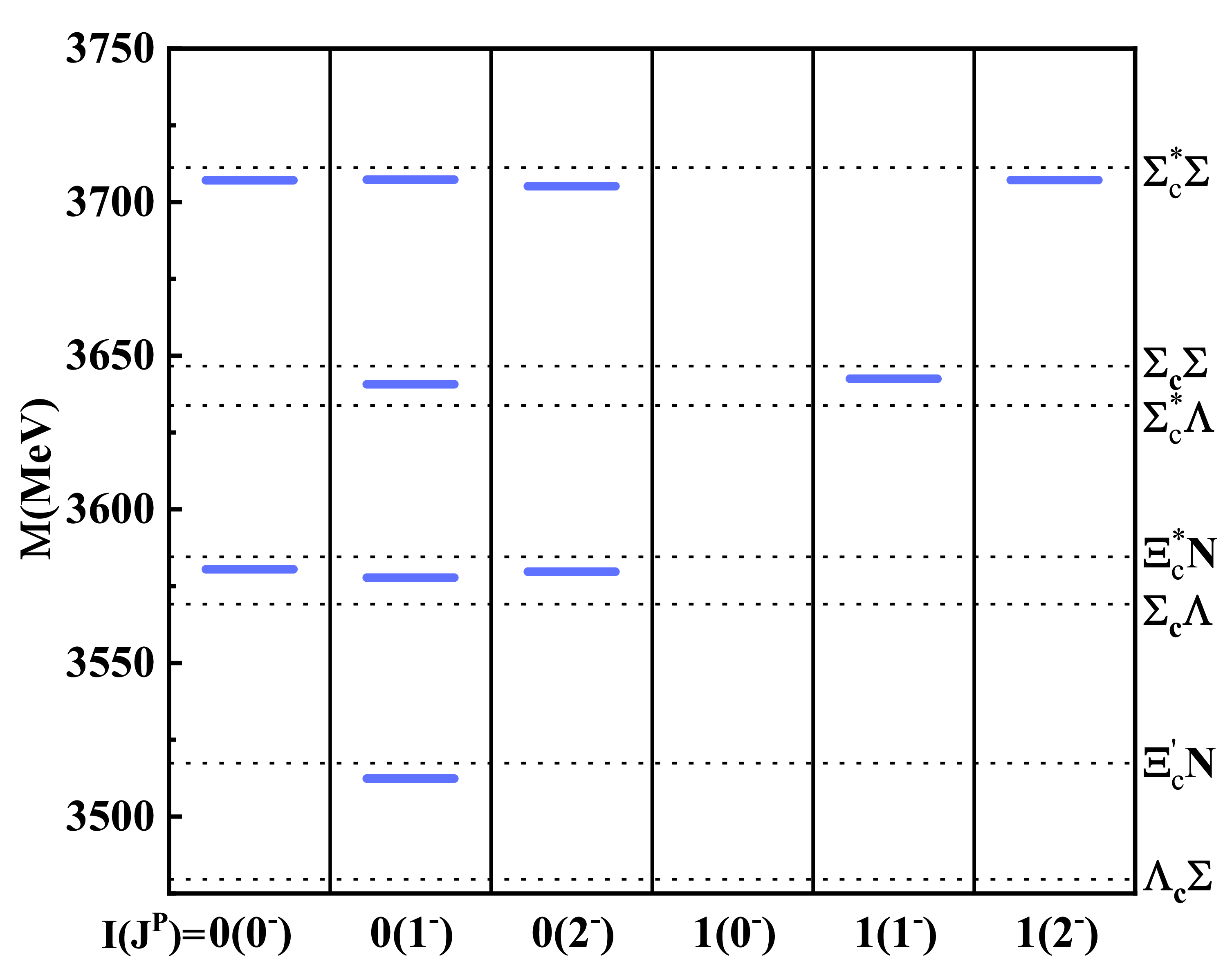}
\caption{Predictions of the possible charm-strange molecular dibaryons with negative parity.}
\label{molecule}
\end{figure}

Specifically, after solving the coupled channel Schr\"{o}dinger equations with the derived OBE effective potentials, a series of potential molecular candidates are obtained as shown in Figure \ref{molecule}, which include the $\Xi_c^{\prime}N$ molecule with $I(J^P)=0(1^-)$, the $\Xi_c^{*}N$ molecules with $0(0^-)$, $0(1^-)$, $0(2^-)$, the $\Sigma_c\Sigma$ molecules with $0(1^-)$ and $1(1^-)$, the $\Sigma_c^*\Sigma$ molecules with $0(0^-)$, $0(1^-)$, $0(2^-)$, and $1(2^-)$. Notably, we also find that coupled channel effects play a significant role in the formation of the $\Sigma_c\Sigma$ molecule with $1(1^-)$.

\begin{figure}[!htbp]
\centering
\includegraphics[width=3.4in]{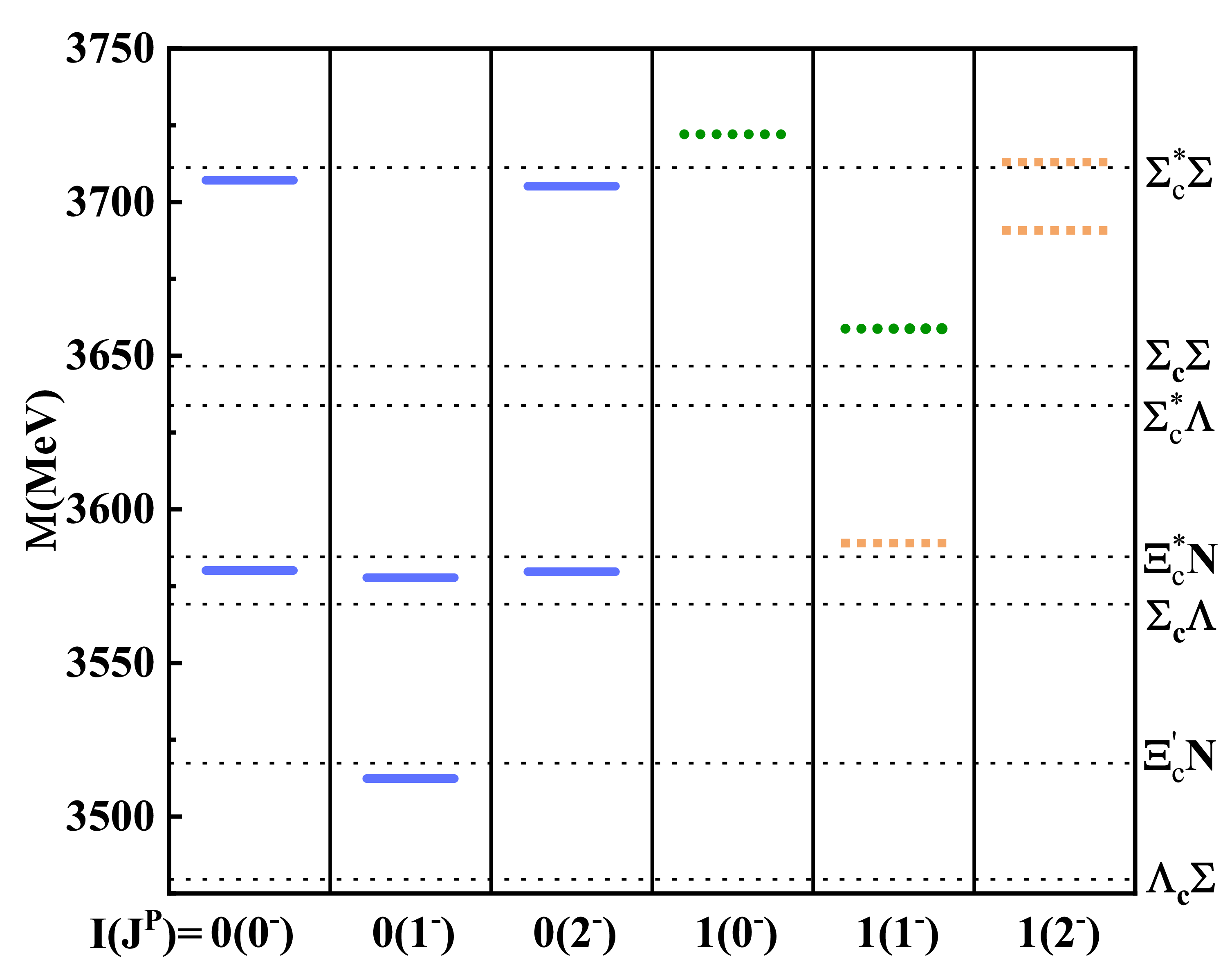}
\caption{A summary of the possible charm-strange resonant dibaryons with negative parity. Here, the blue solid lines, the orange dash lines, and the green dotted lines stand for the independent, shape-type, Feshbach-type resonances, respectively.}
\label{res}
\end{figure}

In addition with predicting potential molecular candidates, our results also offer insights into the possible existences of di-baryon resonances. These resonances could emerge in various coupled interaction channels, such as the $\Xi_cN/\Xi_c^*N$ coupled interactions with $0(0^-, 2^-)$, the $\Xi_c^*N/\Sigma_c^*\Sigma$ coupled interactions with $0(0^-, 2^-)$, the $\Xi_cN/\Xi_c^{\prime}N/\Xi_c^*N$ coupled interactions with $0(1^-)$, the $\Lambda_c\Sigma/\Sigma_c^*\Lambda/\Sigma_c^*\Sigma$ coupled interactions with $1(0^-, 2^-)$, the $\Sigma_c^*\Lambda/\Sigma_c^*\Sigma$ coupled interactions with $1(0^-, 2^-)$, the $\Lambda_c\Sigma/\Sigma_c\Lambda/\Sigma_c\Sigma$ coupled interactions with $1(1^-)$, and the $\Sigma_c^*\Lambda/\Sigma_c\Sigma$ coupled interactions with $1(1^-)$. We subsequently analyze the phase shifts for these coupled-channel systems and find that resonances may indeed exist across all the considered interactions.

Moreover, when comparing the resonant properties to the bound state solutions of the predicted charm-strange molecular dibaryons, we observe that resonances arising from isoscalar coupled channel interactions and from the $\Sigma_c^* \Lambda / \Sigma_c^* \Sigma$ system with $1(2^-)$ cannot be considered as independent states. Instead, their appearances are closely related to the predicted molecular states. Specifically, the $\Xi_c N / \Xi_c^* N$ coupled resonances with $0(0^-, 2^-)$ is close to the $\Xi_c^* N$ molecule with $0(0^-, 2^-)$, the $\Xi_c^* N / \Sigma_c^* \Sigma$ coupled resonance with $0(0^-, 2^-)$ is close to the $\Sigma_c^* \Sigma$ molecule with $0(0^-, 2^-)$, the $\Xi_c N / \Xi_c^{\prime} N / \Xi_c^* N$ coupled resonances with $0(1^-)$ is close to the $\Xi_c^{(\prime,)}$ molecules with $0(1^-)$, and the $\Sigma_c^* \Lambda / \Sigma_c^* \Sigma$ coupled interactions with $1(2^-)$ is close to the $\Sigma_c^* \Sigma$ molecule with $1(2^-)$, all show this close interrelation. In Figure \ref{res}, the relevant resonances are labeled with the same color with those in Figure \ref{molecule}. Additionally, we predict the potential existence of other charm-strange resonant dibaryons, including the $\Sigma_c \Sigma$ shape-type resonance with $1(1^-)$, the $\Sigma_c^* \Sigma$ shape-type resonance with $1(0^-)$, the $\Lambda_c \Sigma / \Sigma_c \Sigma$ coupled Feshbach-type resonance with $1(1^-)$, and the $\Lambda_c \Sigma / \Sigma_c^* \Sigma$ coupled Feshbach-type resonance with $1(0^-, 2^-)$.

Charm-strange dibaryons represent a novel class of matter distinct from traditional mesons and baryons. This study aims to provide comprehensive theoretical predictions regarding the spectral properties of charm-strange dibaryons, derived from both heavy and light baryon-baryon interactions. We hope that future experimental endeavors will be able to confirm the validity of our proposals.

\section*{ACKNOWLEDGMENTS}

This project is supported by the National Natural Science Foundation of China under Grants Nos. 12305139, 12305087. R. C. is supported by the Xiaoxiang Scholars Programme of Hunan Normal University. Q. H. is supported the Start-up Funds of Nanjing Normal University under Grant No.~184080H201B20.


\end{document}